\documentclass[10pt,conference,letterpaper]{IEEEtran}

\usepackage{mathrsfs}
\usepackage{times,amssymb}
\usepackage{amsmath}
\usepackage{subfigure}
\usepackage[font=small]{caption}
\usepackage[dvips]{graphicx}
\usepackage{url}
\newcommand{\subparagraph}{}
\usepackage{amsthm}
\usepackage{multirow}
\usepackage{color}
\usepackage[none]{hyphenat}
\usepackage{titlesec}
\usepackage{float}
\usepackage{epstopdf}
\usepackage{balance}
\usepackage{graphicx,pifont}% http://ctan.org/pkg/{graphicx,pifont}
\let\oldding\ding% Store old \ding in \oldding
\renewcommand{\ding}[2][1]{\scalebox{#1}{\oldding{#2}}}% Scale \oldding via optional argument
\usepackage[dvipsnames]{xcolor}
\usepackage[colorlinks,
            linkcolor=blue,
            anchorcolor=blue,
            urlcolor=blue,
            citecolor=blue]{hyperref}

\usepackage[compress]{cite}
\usepackage[ruled, titlenotnumbered, noend]{algorithm2e}
\usepackage{algorithmicx}
\usepackage[noend]{algpseudocode}          % �㷨����ȥ��end ��
\usepackage{bbm}
\usepackage{subfigure}
\usepackage{bm}
\usepackage{booktabs}

\usepackage{relsize}
\newenvironment{mathalign}{\par\noindent\addtolength{\jot}{-0.3em}\align}{\endalign}

\usepackage{authblk}

\IEEEoverridecommandlockouts
\IEEEpubid{\makebox[\columnwidth]{979-8-3503-9973-8/23/\$31.00 \copyright2023 IEEE \hfill} \hspace{\columnsep}\makebox[\columnwidth]{ }}

\begin{document}
\title{When Computing Power Network Meets Distributed Machine Learning: An Efficient Federated Split Learning Framework}

\author[$\dag$]{Xinjing Yuan}
\author[$\dag$]{Lingjun Pu}
\author[$\S$]{Lei Jiao}
\author[$\ddag$]{Xiaofei Wang}
\author[$\dag$]{Meijuan Yang}
\author[$\dag$]{Jingdong Xu}
\affil[$\dag$]{Institute of Systems and Networks, College of Computer Science, Nankai University, China}
\affil[$\S$]{Department of Computer Science, University of Oregon, USA}
\affil[$\ddag$]{School of Computer Science and Technology, Tianjin University, China}
\affil[ ]{Corresponding author: Lingjun Pu (pulingjun@nankai.edu.cn)}
\maketitle

\allowdisplaybreaks

\begin{abstract}
In this paper, we advocate \textsf{CPN-FedSL}, a novel and flexible Federated Split Learning (FedSL) framework over Computing Power Network (CPN). We build a dedicated model to capture the basic settings and learning characteristics (e.g., training flow, latency and convergence). Based on this model, we introduce Resource Usage Effectiveness (RUE), a novel performance metric integrating training utility with system cost, and formulate a multivariate scheduling problem that maximizes RUE by comprehensively taking client admission, model partition, server selection, routing and bandwidth allocation into account (i.e., mixed-integer fractional programming). We design \textsf{Refinery}, an efficient approach that first linearizes the fractional objective and non-convex constraints, and then solves the transformed problem via a greedy based rounding algorithm in multiple iterations. Extensive evaluations corroborate that \textsf{CPN-FedSL} is superior to the standard and state-of-the-art learning frameworks (e.g., FedAvg and SplitFed), and besides \textsf{Refinery} is lightweight and significantly outperforms its variants and \emph{de facto} heuristic methods under a variety of settings.
\end{abstract}

\section{Introduction}
The significant trend of network and computing convergence in the upcoming Network 2030 \cite{ITU2030, huawei2030} inspires a new concept of Computing Power Network (CPN), also known as Compute First Networking (CFN) \cite{krol2019compute,crowcroft2021compute}. CPN aims to connect diverse computing power in the cloud, on the edge, and across devices to implement on-demand resource scheduling. %In CPN, users only need to customize their service requirements (e.g., QoE and latency), and a dedicated controller will schedule each service request to optimal computing site along optimal path to meet the service requirements.
Recently, ITU-T has launched the first CPN standard \cite{ITU-CPN} to facilitate its proactive development, and many organizations have incorporated CPN into their future research, such as ``Eastern Data and Western Computing" in China and ``Sky Computing" advocated by UC Berkeley \cite{stoica2021cloud}. It can be foreseen that CPN will be a key scenario to boost many emerging services (e.g., extended reality and big data analytics).

As distributed machine learning plays a key role in future large-scale data collection and analytics, we consider it should be one of the killer services in CPN. Recently, integrating federated learning (FL) with split learning (SL), federated split learning (FedSL) is introduced as a promising distributed machine learning paradigm that offers many advantages, such as data privacy, parallel training and lightweight on-device computation requirement \cite{thapa2020splitfed, han2021accelerating, oh2022locfedmix, hongefficient, zhang2023privacy}. However, directly deploying existing FedSL frameworks in CPN is inefficient, since they cannot fully utilize the computing power at different computing sites and overlook the heterogeneity of data clients (e.g., different computing capacity and different data amount). Specifically, they generally adopt only one training server and split the training model in the same manner across all the clients (i.e., only one partition point), which could prolong the global aggregation and consequently damage the model convergence. Therefore, it is of significance to design a flexible and efficient FedSL framework in CPN.

\begin{figure}[tt]
\centering
\setlength{\belowcaptionskip}{-0.3cm}
\includegraphics[width=\linewidth]{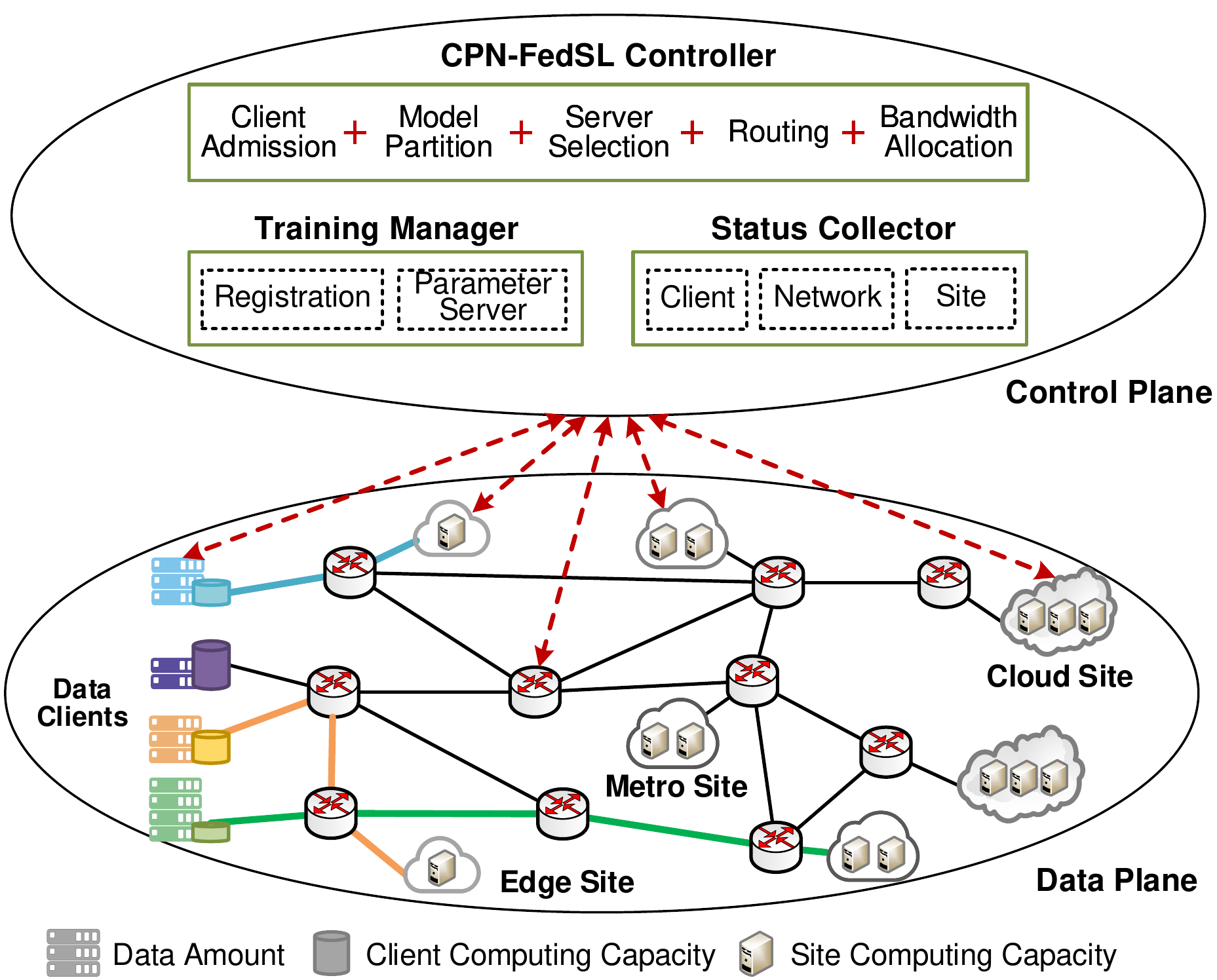}
\caption{The overview of \textsf{CPN-FedSL} (The computing sites include cloud sites, metro sites, edge sites and even mobile devices).}
\label{fig:FedSL}
\end{figure}

In this paper, we advocate \textsf{CPN-FedSL} as shown in Fig. \ref{fig:FedSL}, a novel and flexible federated split learning framework over computing power network, where the registered users (i.e., cooperative data clients) only need to customize their training task (e.g., the training model, the number of training rounds and the expected training accuracy), then a service controller will take over the resource scheduling for federated split learning per training round, and besides a dedicated parameter server will conduct the global model aggregation. Compared with existing FedSL frameworks, \textsf{CPN-FedSL} will assign an exclusive virtualized training server (e.g., in the form of containers) to each client, in terms of distributed computing sites, and it will further independently split the training model for each pair of client and server.

Despite potential benefits, designing efficient \textsf{CPN-FedSL} is challenging, since the service controller requires to tackle the following multivariate scheduling:

\emph{Client Admission}. FedSL intrinsically inherits the characteristics of both FL and SL. As such, choosing more clients per global round in FedSL could speed up the model convergence from the perspective of FL (e.g., FedAvg \cite{li2019convergence} and FedSGD \cite{yuan2020federated}), while leading to more exchanged data and computing workloads at the server side from the perspective of SL \cite{vepakomma2018split, tuli2022splitplace}. Naturally, there is a tradeoff between training utility and system cost, and accordingly the client admission should be considered in order to jointly optimize them.

\emph{Model Partition, Server Selection and Routing}. When the client admission is given, a joint model partition, server selection and routing should be addressed. This is because model partition concurrently determines the amount of computing workloads at the server side and the amount of exchanged data between client and server, which affects the decision-making of server selection and routing. Moreover, all of them have effect on both training utility and system cost. Further complicating this joint decision are the global model aggregation and the bandwidth competition for network links. The former one requires each chosen client and its assigned server should jointly accomplish the training per global round within a given deadline, in order to avoid the cask effect that delays the global aggregation, and the latter one results in the correlation among the routings for different pairs of client and server.

\emph{Bandwidth Allocation}. In the context of CPN, we can control the end-to-end bandwidth for each service \cite{ITU-CPN}. Intuitively, allocating a large amount of bandwidth can facilitate the service performance, while it could lead to a big bandwidth waste. Therefore, on-demand bandwidth allocation is required.

The above intertwined client admission, model partition, server selection, routing and bandwidth allocation (i.e., the multivariate scheduling) is non-trivial and as far as we know has not been explored. Briefly, the previous FL-based studies have extensively investigated the client admission, while they do not touch upon the model partition (see the surveys \cite{lim2020federated,khan2021federated} as references). Although the previous SL-based studies \cite{wu2022split, krouka2021communication, tuli2022splitplace, gao2021evaluation} pay attention to the model partition, they conduct the training process sequentially across the clients (i.e., sequential training), different from the parallel training in \textsf{CPN-FedSL}. Moreover, they consider a single partition point, also different from ours. There are a variety of resource optimization studies on distributed (edge) clouds including service placement, workload distribution and request routing, but none of them fully captures the above multivariate scheduling (see the surveys \cite{mao2017survey, sonkoly2021survey} as references). Despite the resource optimization studies on task offloading, especially the recent neural network partition and offloading for model inference (e.g., \cite{gao2021ocdst, gao2019deep, he2020joint, zhang2021autodidactic}) are similar to ours, most of them optimize a single objective (e.g., the inference latency), while our objective is to jointly optimize training utility and system cost. In addition, they consider single user scenario or neglect the bandwidth competition among multiple users.

In order to efficiently address the above multivariate scheduling for \textsf{CPN-FedSL}, we build a dedicated model that captures the basic settings (e.g., the multivariate scheduling) and learning characteristics (e.g., training flow, latency and convergence) of FedSL over CPN. Then, we model training utility and system cost. On this basis, we introduce Resource Usage Effectiveness (RUE), a novel performance metric that integrates training utility with system cost (i.e., RUE $\triangleq$ Utility$/$Cost). In the end, we formulate a RUE maximization problem for the multivariate scheduling (Sec. \ref{sec2}).

We design \textsf{Refinery}, an efficient approach to address the formulated mixed-integer fractional programming problem. Briefly, \textsf{Refinery} resorts
to the Dinkelbach's transform \cite{dinkelbach1967nonlinear} to linearize the fractional objective, and further decouples the product of control variables
in the non-convex constraints with their interior relationship. Then, it iteratively solves the transformed NP-Hard problem (i.e., a new variant of unsplittable multi-commodity flow problem) via a greedy based rounding algorithm, until the objective converges (Sec. \ref{sec3}).

We consider two kinds of training tasks (i.e., MobileNet and DenseNet) and synthesize different scales of \textsf{CPN-FedSL} in terms of two realistic network topologies (i.e., NSFNET and USNET). Extensive evaluations demonstrate that (1) \textsf{CPN-FedSL} is superior to the standard and state-of-the-art learning frameworks (e.g., FedAvg \cite{li2019convergence} and SplitFed \cite{thapa2020splitfed}); (2) Compared with different variants, \textsf{Refinery} verifies that each part of the multivariate scheduling has noticeable effect on RUE; (3) Despite there is no other approach that jointly optimizes training utility and system cost (e.g., RUE) in terms of the multivariate scheduling considered in this paper, \textsf{Refinery} significantly outperforms several \emph{de facto} heuristic methods under a variety of settings; (4) The proposed greedy based rounding algorithm is reasonably good, i.e., it can empirically reach 65\% \!--\! 80\% of the optimum (Sec. \ref{sec4}).

\section{Framework Model}\label{sec2}

We first provide the basic settings and learning characteristics of \textsf{CPN-FedSL}. Then, we present the model of training utility and system cost, and on this basis we introduce Resource Usage Effectiveness, an integrated performance metric for \textsf{CPN-FedSL}. Finally, we give the problem formulation.

\subsection{Basic Settings and Learning Characteristics}

\begin{figure*}[tt]
\centering
\setlength{\belowcaptionskip}{-0.3cm}
\includegraphics[width=0.85\linewidth]{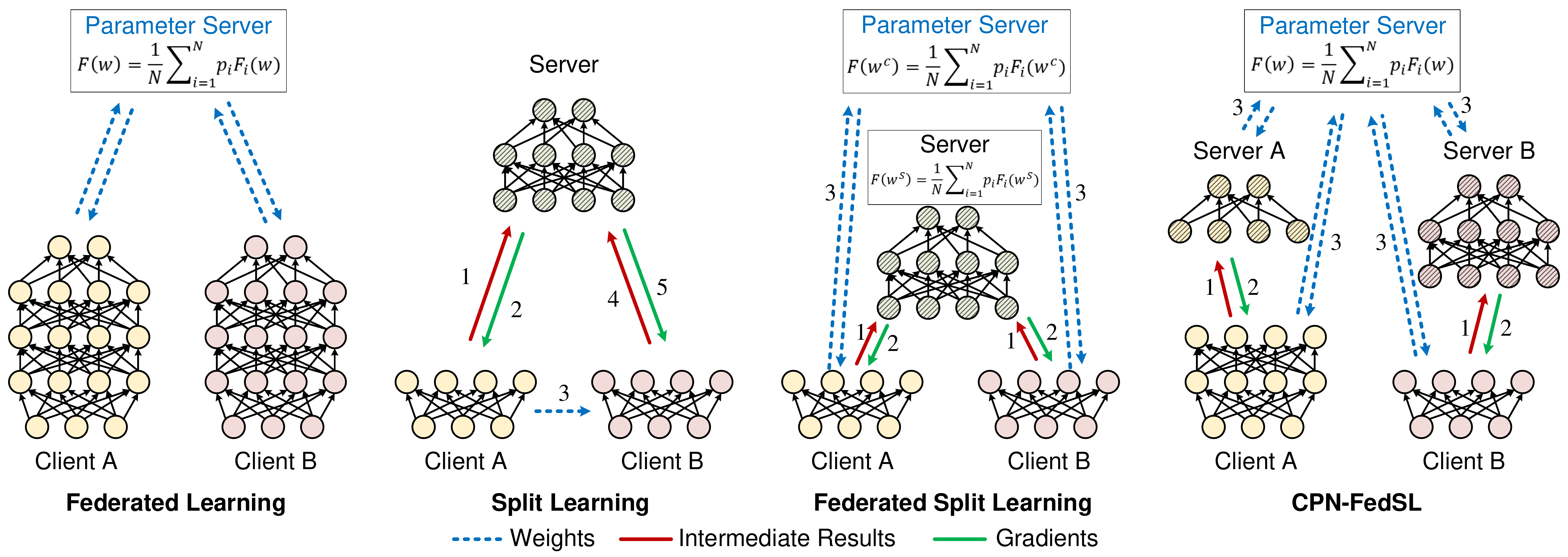}
\caption{Distributed machine learning paradigms (FL: The model training is in the client, and the parameter server calculates the average weights among clients and overrides the local weights \cite{li2019convergence}; SL: The server partition sequentially trains with each of the clients and the client weights are shared with the next training client \cite{vepakomma2018split}; FedSL: The server trains clients' intermediate results (i.e., forward-prorogation of the cut-layer's activation and the labels of data samples) in parallel and the clients' weights are averaged by a parameter server \cite{thapa2020splitfed}; \textsf{CPN-FedSL}: Multiple server and client pairs train simultaneously in an SL manner, and their weights are averaged in an FL manner).}
\label{fig:comparison}
\end{figure*}

\textbf{Network Scenario}. We consider that \textsf{CPN-FedSL} can be viewed as a dedicated network slicing in computing power network for the federated split learning service. As shown in Fig. \ref{fig:FedSL}, the data plane of \textsf{CPN-FedSL} can be modeled by a directed graph $\mathcal{G}\!=\!\{\mathcal{V},{E}\}$, where $\mathcal{V}\triangleq\{\mathcal{V}_C, \mathcal{V}_R, \mathcal{V}_S\}$ refers to the union set of data clients (e.g., users, hospitals and companies), CPN routers (e.g., supporting segment routing) and computing sites (e.g., edge, metro and cloud sites), and ${E}$ refers to the set of physical network links. The control plane of \textsf{CPN-FedSL} can comprehensively monitor the computing and network resources, and efficiently schedule each training request to optimal computing site along optimal path.

In this paper, we simply assume \textsf{CPN-FedSL} will provide an exclusive sub-slicing for each training task\footnote{Similar to many FL, SL and FedSL studies, we will independently treat each training task (i.e, different training tasks do not interfere with each other).}. In this context, we consider each training task is conducted by a set $\mathcal{N}\!\subseteq\!\mathcal{V}_C$ of federated clients with the assistance of a set $\mathcal{M}\!\subseteq\!\mathcal{V}_S$ of computing sites. These $N\!=\!|\mathcal{N}|$ clients and $M\!=\!|\mathcal{M}|$ computing sites are connected by a set $\mathcal{R}\!\subseteq\!\mathcal{V}_R$ of CPN routers, and we consider there are multiple available paths denoted by $\mathcal{L}_{ij}$ for each pair of client $i$ and computing site $j$. Intuitively, each path consists of a series of physical network links.

\textbf{Training Setting}. As shown in Fig. \ref{fig:comparison}, similar to many FL and FedSL frameworks, we consider the control plane of \textsf{CPN-FedSL} which is physically located in a given computing site will deploy a parameter server in the same computing site for each training task. This parameter server takes charge of global model aggregation and aims to derive the optimal training model $\bm{\mathrm{w}}^*$ (i.e, model parameter vector) that minimizes the loss function $F(\bm{\mathrm{w}})\!=\!\frac{1}{N}\sum_{i=1}^Np_iF_i(\bm{\mathrm{w}})$ within $T$ global rounds. Note that $F_i(\bm{\mathrm{w}})$ refers to the empirical loss function of client $i$ defined as $F_i(\bm{\mathrm{w}})\!=\!\frac{1}{|D_i|}\sum_{d\in D_i}f(d;\bm{\mathrm{w}})$, where $D_i$ is the local dataset of client $i$, and $f(d;\bm{\mathrm{w}})$ is the loss computed by the current model $\bm{\mathrm{w}}$ and the data sample $d$. Besides, $p_i$ refers to the weight of client $i$ with $\sum_{i=1}^Np_i\!=\!1$. %For example, $p_i$ can be set to $|D_i|/\sum_{i=1}^N|D_i|$.

Different from previous SL and FedSL frameworks which adopt only one training server and split the training model $\bm{\mathrm{w}}$ into the same client-side module $\bm{\mathrm{w}}^C$ and server-side module $\bm{\mathrm{w}}^S$ across all the clients (i.e., only one partition point), we consider each client $i$ is assigned an exclusive virtualized training server (e.g., container) in a computing site $j$, and the training model for each pair of client and server is independently partitioned into two modules as $\bm{\mathrm{w}}_{t}\!=\!\big[\bm{\mathrm{w}}^C_{it}(k),\bm{\mathrm{w}}^S_{it}(k)\big], \forall i\!\in\!\mathcal{N}$ per global round $t\!\in\!\{1,2,\dots,T\}$, where $k\!\in\!\{1,2,\dots,K_{\bm{\mathrm{w}}}\}$ indicates the partition point (i.e., the cut-layer), and $K_{\bm{\mathrm{w}}}$ is the total number of layers of training model $\bm{\mathrm{w}}$. Note that $k\!=\!K_{\bm{\mathrm{w}}}$ refers to client local training, and we do not allow $k=0$ (i.e., data samples forwarding and training only at the server side) due to data privacy issue.

\textbf{Basic Notations}. We introduce a series of notations to facilitate our following descriptions. Specifically, we denote by $c_{it}$ the computing capacity of client $i$ in round $t$, which remains the same within each global round while could change across different rounds. We consider each computing site $j$ can provide a number of virtualized training servers (e.g., containers), and we denote by $w_j$ the computing capacity of each training server and by $\Omega_j$ the number of available training servers. In other words, the training servers are homogeneous in each computing site with the help of advanced virtualization techniques such as Kubernetes and Nvidia vGPU. As to a training model, given the partition point $k$, we denote by $q_k^C$ the computing density (e.g., FLOPS) of the client-side module, by $q_k^S$ the computing density of the server-side module and by $s_k$ the amount of exchanged data between client and server per training batch. As to a network link $e\!\in\!E$, we denote by $B_e$ the bandwidth capacity.

\textbf{Remark}. Given a training task, the \textsf{CPN-FedSL} controller as shown in Fig. \ref{fig:FedSL} can derive $q_k^C$, $q_k^S$ and $s_k$ in an offline manner, obtain $p_i$ during client registration, collect $c_{it}$ periodically from client uploading (i.e., status monitoring), and set values for $w_j$, $\Omega_j$ and $B_e$ as the sub-slicing for that task.

\textbf{Multivariate Scheduling}. As to \emph{client admission}, we introduce a binary control variable $z_{it}\!\in\!\{0,1\}$ to indicate if client $i$ is chosen in round $t$. As to \emph{model partition}, \emph{server selection} and \emph{routing}, we introduce a binary control variable $x_{ijt}^{kl}\!\in\!\{0,1\}$ to indicate if client $i$ is assigned a training server in computing site $j$ connected by the path\footnote{We consider the general single-path routing (i.e., unsplittable flow), and we can exploit the segment routing technique to implement it in practice \cite{he2022enabling}.} $l\!\in\!\mathcal{L}_{ij}$ when the partition point is $k$ in round $t$. As to bandwidth allocation, we introduce a control variable $y_{ijt}^{k}\!\in\!\mathbb{R}^+$ to indicate the allocated end-to-end bandwidth between client $i$ and computing site $j$ when the partition point is $k$ in round $t$. Intuitively, we have the following constraints:
\begin{mathalign}
&\mathsmaller{\sum\nolimits_j\sum\nolimits_k\sum\nolimits_l x_{ijt}^{kl}=z_{it},}\label{const1}
\end{mathalign}
\begin{mathalign}
&\mathsmaller{\sum\nolimits_i\sum\nolimits_k\sum\nolimits_l x_{ijt}^{kl}\le \Omega_j,}\label{const2}
\end{mathalign}
\begin{mathalign}
\mathsmaller{\sum\nolimits_i\sum\nolimits_j\sum\nolimits_k\sum\nolimits_ly_{ijt}^kx_{ijt}^{kl}\mathbbm{1}_{ij}^{l(e)}\le B_{e}}.\label{const3}
\end{mathalign}
The constraint (\ref{const1}) indicates that an exclusive training server in a computing site should be assigned to each chosen client $i$, the constraint (\ref{const2}) indicates that each computing site $j$ can serve at most $\Omega_j$ clients due to the number of available training servers, and the constraint (\ref{const3}) refers to the link bandwidth capacity constraint where $\mathbbm{1}_{ij}^{l(e)}$ is a binary indicator that indicates whether network link $e$ is on path $l\!\in\!\mathcal{L}_{ij}$ between client $i$ and computing site $j$.

\textbf{Training Flow}. In terms of the preceding discussions, we next present the training flow for each task, with the initial model $\bm{\mathrm{w}}_0$, the number of epochs $\mathcal{E}$ and the batch size $H$.

\emph{Step 1 (Multivariate Scheduling)}: In the beginning of each round $t$, the \textsf{CPN-FedSL} controller first collects the status%\footnote{As mentioned in the \textbf{Remark} on page 3, the controller can obtain other system parameters in advance. In practice, it can also support time-varying bandwidth capacity of network links (i.e., $B_{et}$) and computing capacity of computing sites (i.e., $w_{jt}$) via periodical network and site status monitoring.} 
of data clients (i.e., $c_{it}$) and strategically chooses a set $\mathcal{A}_t$ of clients, assigns an exclusive training server in a computing site to each client $i\in \mathcal{A}_t$, decides the model partition point $k$, the end-to-end routing path and its allocated end-to-end bandwidth for each pair of client and server.

\emph{Step 2 (Model Download)}: The parameter server exploits multicast to send the previous aggregated model $\bm{\mathrm{w}}_{t}$ to the chosen clients in $\mathcal{A}_t$ and their corresponding training servers. Then, each pair of client and server can obtain the corresponding client-side module $\bm{\mathrm{w}}^C_{it}(k)$ and server-side module $\bm{\mathrm{w}}^S_{it}(k)$ in terms of $\bm{\mathrm{w}}_{t}$ and the specified partition point $k$.

\emph{Step 3 (Split Model Training)}: According to the batch size, each client $i\in \mathcal{A}_t$ has ${|D_i|}/{H}$ batches. For each batch per epoch, client $i$ executes forward-propagation (FP) with the current batch, sends the FP of the last layer's activation of client-side module $\bm{\mathrm{w}}^C_{it}(k)$ associated with the data labels in the current batch to the assigned server. The server proceeds FP, computes the loss and generates the gradient to update the server-side module (i.e., $\bm{\mathrm{w}}^S_{it}(k)\!\to\!\bm{\mathrm{w}}^S_{it+1}(k)$). Then, the server sends the back-propagation of the activation's (i.e., the cut-layer's) gradient back, with which client $i$ can update the client-side module. Note that the above procedure iteratively operates $\mathcal{E}\!\times\!{|D_i|}/{H}$ (i.e., epoch$\times$batch) times per round $t$.

\emph{Step 4 (Model Aggregation)}: Each client $i\in \mathcal{A}_t$ uploads a synthetic model $\bm{\mathrm{w}}_{it+1}'\!=\!\big[\bm{\mathrm{w}}^C_{it+1}(k),\bm{0}\big]$, and its assigned server uploads the counterpart model $\bm{\mathrm{w}}_{it+1}''\!=\!\big[\bm{0},\bm{\mathrm{w}}^S_{it+1}(k)\big]$ to the parameter server (i.e., $\bm{\mathrm{w}}_{it+1}\!=\!\bm{\mathrm{w}}_{it+1}'\!+\!\bm{\mathrm{w}}_{it+1}''$) for model aggregation (e.g., FedAvg). Then, $t=t+1$ and go to Step 1.

\textbf{Training Latency Analysis}. Given the multivariate scheduling (i.e., $x_{ijt}^{kl},y_{ijt}^k, z_{it}$ are decided), we denote by $\tau_{it}^0$ the time when the controller sends scheduling decision and previous aggregated model to the chosen client $i$ (i.e., latency in Step 1 and Step 2), which can be expressed by
\begin{mathalign}
&\tau_{it}^0=(\delta+|\bm{\mathrm{w}}_{t}|)z_{it}/b_{it},\notag
\end{mathalign}
where $\delta$ and $|\bm{\mathrm{w}}_{t}|$ respectively refer to the size of scheduling decision and aggregated model, and $b_{it}$ refers to the end-to-end bandwidth between client $i$ and the parameter server. Note that $b_{it}$ is not a control variable but can be estimated by historical data samples (i.e., bandwidth prediction).

To proceed, we denote by $\tau_{it}^1$ the training latency of each pair of client and server in Step 3, which can be expressed by
\begin{mathalign}
&{\tau_{it}^1=\frac{\mathcal{E}|D_i|}{H}\mathsmaller{\sum\nolimits_j\sum\nolimits_k\sum\nolimits_l}\Big[\frac{q_k^C}{c_{it}}+\frac{q_k^S}{w_{j}}+\frac{s_k}{y_{ijt}^k}\Big]x_{ijt}^{kl},}\notag
\end{mathalign}
where the first term is the total computing time at the client-side, the second term is the total computing time at the server-side and the third term is the total data (i.e., activation and gradient) transmission time between client and server.

At last, we denote by $\tau_{it}^2=(|\bm{\mathrm{w}}'_{it+1}|+\delta')z_{it}/b_{it}$ the time when client $i$ uploads its synthetic model and its predicted computing capacity whose sizes are respectively $|\bm{\mathrm{w}}'_{it+1}|$ and $\delta'$ to the controller. Note that we do not need to incorporate the communication time (i.e., model download and upload) between the controller and the assigned servers. For example, the server model download in Step 2 must have finished before client sends the intermediate results to server in Step 3.

To avoid the cask effect that several slow client-server pairs delay the model aggregation per global round, we have the following training latency constraint for each client-server pair ($\Delta$ refers to the time length of global round):
\begin{mathalign}
&\tau^0_{it}+\tau^1_{it}+\tau^2_{it}\le \Delta.\label{const4}
\end{mathalign}

\textbf{Training Convergence Analysis}. Besides the common assumptions such as the global and local function $F$ and $F_i$ satisfy \emph{L-smooth}, \emph{bounded local variance} and \emph{bounded gradient} \cite{li2019convergence, yuan2020federated, vepakomma2018split, tuli2022splitplace, thapa2020splitfed, han2021accelerating, oh2022locfedmix, hongefficient, zhang2023privacy}, we also assume that: \emph{the chosen set $\mathcal{A}_t$ at least contains $\mathcal{K}$ clients randomly selected per global round according to the sampling probabilities $p_1, p_2, \dots, p_N$}. In this context, if the global and local function $F$ and $F_i$ are strongly convex, we can derive the following convergence bound:
\begin{mathalign}
&{\mathbb{E}\big[F(\bm{\mathrm{w}}_T)\big]\!-\!F(\bm{\mathrm{w}}^*)\!=\!\mathsmaller{\mathcal{O}\big(\frac{1}{T}+\frac{1}{\mathcal{K}T}\big)}}.\label{const5}
\end{mathalign}
Alternatively, if they are non-convex, we can derive the following convergence bound:
\begin{mathalign}
&\mathsmaller{\frac{1}{T}\sum\nolimits_{t=1}^T\mathbb{E}\big[||\nabla F(\bm{\mathrm{w}}_t)||^2_2\big]\!=\!{\mathcal{O}\big(\frac{1}{T}+\frac{1}{\sqrt{\mathcal{K}T}}\big)}}.\label{const6}
\end{mathalign}

To facilitate our following discussions, we only highlight the relevant components regarding the multivariate scheduling in the above convergence bound. Note that we can prove the bound for strongly convex function in (\ref{const5}) with a similar procedure in \cite{li2019convergence} which however does not consider the theoretical analysis for the non-convex case. We complement it with (\ref{const6}), and the detailed proof are provided in our online technical report \cite{report} for clarity.

%We should emphasize that the preceding training flow associated with the corresponding training latency and convergence analysis is a reasonable but not the only choice in \textsf{CPN-FedSL}, such as exploiting other model aggregation methods rather than FedAvg.

\subsection{Training Utility and System Cost}

\textbf{Training Utility}. In terms of the above convergence analysis (i.e., the assumption on the chosen set $\mathcal{A}_t$ and the derived convergence bound), we can lower the bound by proportionally increasing the value of $\mathcal{K}$. To this end, we introduce the following fairness-aware client admission metric to approximately indicate the training utility per global round:
\begin{mathalign}
&\mathsmaller{U_t=\sum\nolimits_ip_iz_{it}\!+\!\lambda\sum\nolimits_i Q_i(t)z_{it}=\sum\nolimits_i\big(p_i\!+\!\lambda Q_i(t)\big)z_{it},}\notag
\end{mathalign}
where $Q_i(t+1)=Q_i(t)-z_{it}+p_i$ with $Q(0)=0$. Note that a large value of $U_t$ refers to a high training utility, which can be interpreted as follows: (1) the term $\sum\nolimits_ip_iz_{it}$ emphasizes the clients with a higher weight $p$ (i.e., the larger size of local dataset), which contributes to increasing the number of training data samples; (2) the term $\sum\nolimits_iQ_i(t)z_{it}$ enables the average number of times each client is chosen consists with the given sampling probability throughout the training (i.e., fairness), which is inspired from Lyapunov virtual queues \cite{neely2010stochastic} and queuing theory (i.e., the average service rate $\sum\nolimits_{t=1}^Tz_{it}/T$ is no less than the average arrival rate $p_i$ if the virtual queue $Q_i(t)$ is stable). Intuitively, if client $i$ has not been chosen in the recent global rounds, then its weight $Q_i(t)$ will accumulate continuously, and consequently it is more likely to be chosen in the next few rounds. Besides, we allow the weight $Q_i(t), \forall i$ to take negative values so as to avoid the case that several clients are frequently chosen. The parameter $\lambda$ is used to balance those two terms.

\textbf{System Cost}. In the context of Computing Power Network, the service controller should efficiently exploit the computing resources and the network resources. To this end, we introduce the following system cost model per round in \textsf{CPN-FedSL}:
\begin{mathalign}
&\mathsmaller{C_t=\Big\{\sum\nolimits_j\alpha_{jt}\sum\nolimits_i\sum\nolimits_k\sum\nolimits_lx_{ijt}^{kl}}\notag\\
& \mathsmaller{\quad \ +\sum\nolimits_e\beta_{et}\sum\nolimits_i\sum\nolimits_j\sum\nolimits_k\sum\nolimits_ly_{ijt}^kx_{ijt}^{kl}\mathbbm{1}_{ij}^{l(e)}}\notag\\
& \mathsmaller{\quad \ +\sum\nolimits_i\sum\nolimits_j\sum\nolimits_k\sum\nolimits_l\big[\gamma_{it}+\gamma'_{jt}\big]x_{ijt}^{kl}\Big\}\Delta},\notag
\end{mathalign}
where $\sum\nolimits_i\sum\nolimits_k\sum\nolimits_lx_{ijt}^{kl}$ indicates the number of occupied training servers in computing site $j$, $\sum\nolimits_i\sum\nolimits_j\sum\nolimits_k\sum\nolimits_ly_{ijt}^kx_{ijt}^{kl}\mathbbm{1}_{ij}^{l(e)}$ indicates the amount of occupied bandwidth on network link $e$, $\alpha_{jt}$ refers to the unit server cost in computing site $j$, and $\beta_{et}$ refers to the unit bandwidth cost on network link $e$ per round. Besides the above cost regarding the split model training in Step 3, we also introduce $\gamma_{it}+\gamma'_{jt}$ to indicate the communication cost for status collection, model download and upload in Step 1, 2 and 4. Note that $\Delta$ refers to the time length of global round as mentioned before.

\textbf{Resource Usage Effectiveness}. We attempt to jointly optimize training utility and system cost in \textsf{CPN-FedSL}. To achieve this goal, we introduce Resource Usage Effectiveness (RUE), a novel performance metric to integrate them, which indicates how efficiently CPN uses resources for FedSL. We formally present it as
\begin{mathalign}
&\mathsmaller{\text{RUE}=\frac{1}{T}\sum\nolimits_{t=1}^T{U_t}/{C_t}},\notag
\end{mathalign}
the average training utility achieved by unit system cost per round. In this context, our objective is to maximize RUE.

\subsection{Problem Formulation}

In terms of the preceding discussions, we next formulate a multivariate scheduling problem that maximizes RUE by comprehensively taking client admission, model partition, server selection, routing and bandwidth allocation into consideration:
\begin{align*}
& \mathrm{max}  & & \text{RUE}\notag\\
& \mathrm{s.\,t.} & &  \text{(\ref{const1}), \ (\ref{const2}), \ (\ref{const3}), \ (\ref{const4}),}  \notag \\
& \mathrm{var} & &  z_{it}\in\{0,1\}, \ x_{ijt}^{kl}\in\{0,1\}, \ y_{ijt}^k\in\mathbb{R}^+.
\end{align*}
To facilitate the challenge analysis and algorithm design, we make the following simplification. First, we omit the
round index $t$, since the multivariate scheduling is independent across different global rounds (Note that $Q_i$ is not a control variable but an input parameter updated per round). Second, we substitute $z_i$ with $\sum\nolimits_j\sum\nolimits_k\sum\nolimits_lx_{ij}^{kl}$ in terms of constraint (\ref{const1}). In this context, the formulated problem becomes

\begin{align*}
& \mathrm{max}  & & \mathcal{P}_0=\mathsmaller{{\sum\nolimits_i\sum\nolimits_j\sum\nolimits_k\sum\nolimits_l \big(p_i\!+\!\lambda Q_i\big)x_{ij}^{kl}}\over{\sum\nolimits_i\sum\nolimits_j\sum\nolimits_k\sum\nolimits_l\big(\alpha'_{ij}x_{ij}^{kl}+\sum\nolimits_e\beta'_{e}\mathbbm{1}_{ij}^{l(e)}y_{ij}^kx_{ij}^{kl}\big)}}\notag\\
& \mathrm{s.\,t.} & & \mathcal{C}_1:\ \ \mathsmaller{\sum\nolimits_j\sum\nolimits_k\sum\nolimits_l x_{ij}^{kl}\le 1,}\\
& & & \mathcal{C}_2: \ \ \mathsmaller{\sum\nolimits_i\sum\nolimits_k\sum\nolimits_l x_{ij}^{kl}\le \Omega_j,}\\
& & & \mathcal{C}_3:\ \ \mathsmaller{\sum\nolimits_i\sum\nolimits_j\sum\nolimits_k\sum\nolimits_ly_{ij}^kx_{ij}^{kl}\mathbbm{1}_{ij}^{l(e)}\le B_{e}},\\
& & & \mathcal{C}_4:\ \ \mathsmaller{\sum\nolimits_j\sum\nolimits_k\sum\nolimits_l}\Big[\mu_{ij}^k+{s'_k}/{y_{ij}^k}\Big]x_{ij}^{kl}\le\Delta,\\
& \mathrm{var} & &  x_{ij}^{kl}\in\{0,1\}, \ y_{ij}^k\in\mathbb{R}^+,
\end{align*}
where $\alpha'_{ij}\triangleq(\alpha_{j}\!+\!\gamma_i\!+\!\gamma'_j)\Delta$, $\beta'_{e}\triangleq\beta_{e}\Delta$, $\mu_{ij}^k\triangleq (\delta\!+\!\delta'\!+\!2|\bm{\mathrm{w}}|)/b_{it}\!+\!\mathcal{E}|D_i|q_k^C/(Hc_{it})\!+\!\mathcal{E}|D_i|q_k^S/(Hw_j)$ and $s'_k\triangleq \mathcal{E}|D_i|s_k/H$.

\section{Multivariate Scheduling Algorithm}\label{sec3}

\textbf{Algorithmic Challenges}. To tackle the problem $\mathcal{P}_0$, we should address two main challenges: First, $\mathcal{P}_0$ belongs to the mixed-integer fractional programming, which is generally intractable. Second, $\mathcal{P}_0$ incorporates non-convex constraints $\mathcal{C}_3$ and $\mathcal{C}_4$, which further
complicates the algorithm design. To this end, our basic idea is to first linearize the fractional objective and non-convex constraints, and then develop an efficient algorithm to solve the linearized mixed-integer problem.

\vspace{-0.3cm}
\begin{figure}[thb]
\centering
\setlength{\belowcaptionskip}{-0.3cm}
\includegraphics[width=0.9\linewidth]{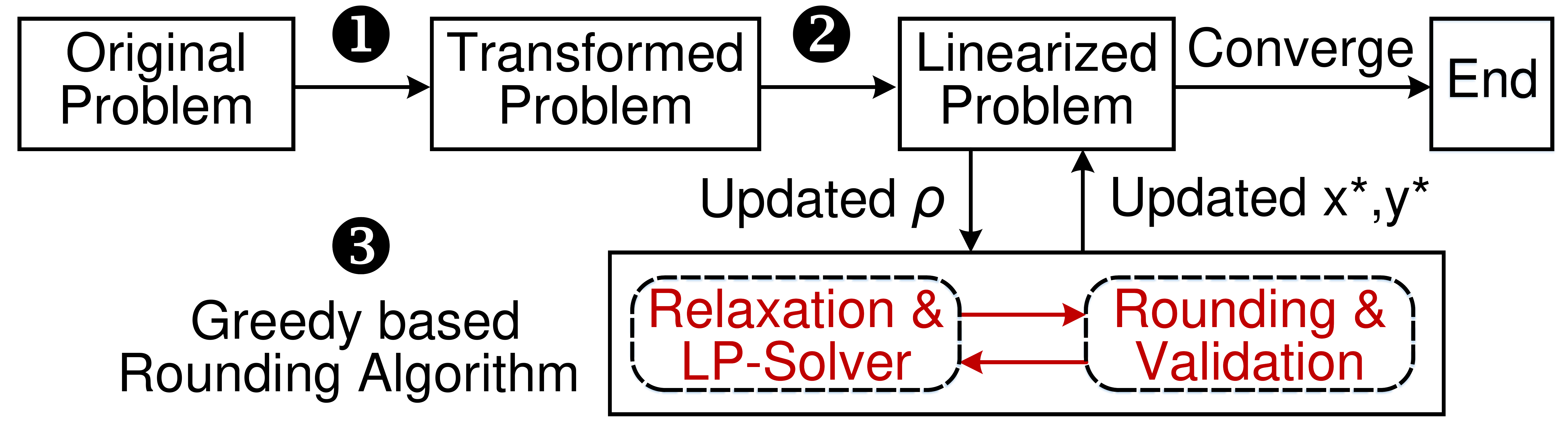}
\caption{Overview of \textsf{Refinery}.}
\label{fig:BasicFlow}
\end{figure}

\textbf{Algorithm Design}. We design \textsf{Refinery} as illustrated in Fig. \ref{fig:BasicFlow}, an efficient approach to solve the problem $\mathcal{P}_0$.

{Step \ding{182}}: We resort to the Dinkelbach's transform \cite{dinkelbach1967nonlinear} to linearize the fractional objective. Specifically, if we treat the objective of the original problem as $\mathcal{P}_0(\bm{x},\bm{y})=\Gamma(\bm{x},\bm{y})/\Psi(\bm{x},\bm{y})$, then we can reformulate it as $\mathcal{P}_0(\bm{x},\bm{y},\rho)=\Gamma(\bm{x},\bm{y})-\rho\Psi(\bm{x},\bm{y})$, where $\rho$ is a parametric variable (initially it is 0). In this context, we can iteratively solve the problem $\mathcal{P}_0$ with the parametric objective $\mathcal{P}_0(\bm{x},\bm{y},\rho)$, the same original constraints and the updated $\rho=\Gamma(\bm{x}^*,\bm{y}^*)/\Psi(\bm{x}^*,\bm{y}^*)$ until the value of the parametric objective $\mathcal{P}_0(\bm{x},\bm{y},\rho)$ converges (i.e., its difference between two continuous iterations is within a given tolerance). At that time, the derived optimal solution (i.e., $\{\bm{x}^*,\bm{y}^*\}$) in terms of the parametric objective is also the optimal solution of the original problem. With this principle, our purpose turns to seeking the optimal solution of the problem $\mathcal{P}_0$ with the parametric objective $\mathcal{P}_0(\bm{x},\bm{y},\rho)$ (i.e., the fractional objective has been linearized).

Step \ding{183}: We linearize the non-convex constraints $\mathcal{C}_3$ and $\mathcal{C}_4$, since they both involve the product of control variables. In term of constraint $\mathcal{C}_4$, if we assume the training of a pair of client $i$ and server $j$ given the partition point $k$ is accomplished exactly at the given deadline, then the control variable $y_{ij}^k$ will be bounded by a constant value $\varphi_{ij}^k$ defined as
\begin{mathalign}
&{\varphi_{ij}^k={s'_k}/{(\Delta-\mu_{ij}^k)}}. \label{const7}
\end{mathalign}

\textbf{Remark}. For each chosen client $i$, given the assigned server $j$ and the partition point $k$, the corresponding end-to-end bandwidth should be allocated at least $\varphi_{ij}^k$ (i.e., $y_{ij}^k\ge\varphi_{ij}^k$). Besides, let the value of $y_{ij}^k$ down to $\varphi_{ij}^k$ has no impact on the optimal solution in terms of constraint $\mathcal{C}_3$ and $\mathcal{C}_4$, since the occupied bandwidth could reduce on some network links given a lower end-to-end bandwidth requirement (i.e., $y_{ij}^k$).

In this context, we can integrate constraint $\mathcal{C}_3$ with $\mathcal{C}_4$ as
\begin{mathalign}
\mathcal{C}_3':\ \ \mathsmaller{\sum\nolimits_i\sum\nolimits_j\sum\nolimits_k\sum\nolimits_l\varphi_{ij}^kx_{ij}^{kl}\mathbbm{1}_{ij}^{l(e)}\le B_{e}}\notag,
\end{mathalign}
and meanwhile modify the corresponding component in the objective function (i.e., $y_{ij}^k=\varphi_{ij}^k$). To sum up, the non-convex constraints have been linearized, and
consequently the original problem becomes a linearized problem with a simplified parametric objective $\mathcal{P}_0(\bm{x},\rho)$.

Moreover, in terms of constraint $\mathcal{C}_3'$, we can have the following theorem to derive the optimal partition point $k^*$ given a pair of client $i$ and server $j$.

\textbf{Theorem 1}. \emph{Given a pair of client $i$ and server $j$, the optimal partition point $k^*$ satisfies $k^*=\arg\min_k \varphi_{ij}^k$ for each $k$ whose $\varphi_{ij}^k$ derived by (\ref{const7}) is positive.}

We can easily prove it by contradiction. Briefly, suppose the optimal partition point is $k'$ whose $\varphi_{ij}^{k'}$ is larger than $\varphi_{ij}^{k^*}$. Then, we can make $\varphi_{ij}^{k'}$ down to $\varphi_{ij}^{k^*}$, which will not violate constraint $\mathcal{C}_3'$ while increasing the value of objective function.

From Theorem 1, we can also derive the optimal end-to-end bandwidth allocation given a pair of client $i$ and server $j$.

\textbf{Corollary 1}. \emph{Given a pair of client $i$ and server $j$, the optimal end-to-end bandwidth allocation $y_{ij}^k$ will be $\varphi_{ij}^{k^*}$.}

According to Theorem 1, we introduce $\varphi_{ij}^*=\min\{\varphi_{ij}^k \lvert \varphi_{ij}^k>0\}$ and we can further simplify the optimization problem as
\begin{align*}
& \mathrm{max}  & & \mathcal{P}_1=\mathsmaller{{\sum\nolimits_i\sum\nolimits_j\sum\nolimits_l \big[p_i\!+\!\lambda Q_i\!-\!\rho(\alpha_{ij}'\!+\!\sum\nolimits_e\beta_{e}'\mathbbm{1}_{ij}^{l(e)}\varphi_{ij}^*)\big]\theta_{ij}^{l}}}\notag\\
& \mathrm{s.\,t.} & & \mathsmaller{\sum\nolimits_j\sum\nolimits_l \theta_{ij}^{l}\le 1,}\\
& & & \mathsmaller{\sum\nolimits_i\sum\nolimits_l \theta_{ij}^{l}\le \Omega_j,}\\
& & & \mathsmaller{\sum\nolimits_i\sum\nolimits_j\sum\nolimits_l\varphi_{ij}^*\theta_{ij}^{l}\mathbbm{1}_{ij}^{l(e)}\le B_{e}},\\
& \mathrm{var} & &  \theta_{ij}^{l}\in\{0,1\}.
\end{align*}

In the following, our purpose turns to seeking the optimal solution of the problem $\mathcal{P}_1$, with which we can derive the corresponding optimal $\{x_{ij}^{kl}\}$ in terms of Theorem 1 (i.e., $\theta_{ij}^{l}+k^*\to x_{ij}^{kl}$) and the optimal $\{y_{ij}^{k}\}$ in terms of Corollary 1.

\textbf{Remark}. The problem $\mathcal{P}_1$ is a new variant of the Unsplittable Multi-commodity Flow Problem (UMFP) with the undecided flow destinations (i.e., undecided client-server pairs) and additional hard capacity constraints (i.e., $\Omega_j$). We can easily prove $\mathcal{P}_1$ is NP-Hard by degrading it to the standard UMFP. Note that if we loosen the capacity constraint of computing sites (i.e., removing the second constraint), then we can resort to the probability theory and Chernoff-type bound to design a tree-based approximation algorithm \cite{chakrabarti2007approximation, raghavan1988probabilistic}. Alteratively,
if we loosen the capacity constraint of network links (i.e., removing the third constraint), then we can optimally solve the problem due to the totally unimodular of its constraints. To the best of our knowledge, there is no existing algorithms that can approximately solve the problem $\mathcal{P}_1$ in polynomial time.

Step \ding{184}: We will develop a greedy based heuristic algorithm to solve the problem $\mathcal{P}_1$ via relaxation, rounding and validation as presented in Alg. \ref{Alg1}.

\begin{algorithm}[tt]
\small
\SetAlgoNoLine
\caption{\small{Greedy based Rounding Algorithm}}\label{Alg1}
\KwIn{\ \ $\mathcal{A}_{acc}$, $\mathcal{A}_{rej}$, $\mathcal{A}_{cur}$, $\hat{\Theta}$}
\KwOut{$\hat{\Theta}$}
{
Initially $\mathcal{A}_{acc}, \mathcal{A}_{rej}, \hat{\Theta}=\emptyset, \mathcal{A}_{cur}\!=\!\mathcal{N}$;\\
%$\triangleright$ \ \emph{Local training at the client side};\\
\textbf{While} {$\mathcal{A}_{cur}\neq\emptyset$} \textbf{do}\\{
\quad \ \ Solve relaxed problem to derive the fractional solution $\bar{\bm{\theta}}$;\\
\quad \ \ Sort $\bar{\bm{\theta}}$ in the descending order of value $\omega_{ij}^l\bar{\theta}_{ij}^l$, where $\omega_{ij}^l$ \\
\quad \ \ refers to $p_i\!+\!\lambda Q_i\!-\!\rho(\alpha_{ij}'\!+\!\sum\nolimits_e\beta_{e}'\mathbbm{1}_{ij}^{l(e)}\varphi_{ij}^*)$ in the objective;\\
\quad \ \ Denoted by $\bar{\Theta}$ the ordered list of $\bar{\bm{\theta}}$;\\
\quad \ \ \textbf{For} {each $\bar{\theta}_{i^*j}^l\in\bar{\Theta}$ in order} \textbf{do}\\{
\quad \ \ \quad Round $\bar{\theta}_{i^*j}^l$ up to 1 and the rest $\{\bar{\theta}_{i^*j'}^{l'}\}$ involving $i^*$ to 0;\\
\quad \ \ \quad Invoke SMT for $\mathcal{A}_{acc}\cup\{i^*\}$ under three constraints;\\
\quad \ \ \quad \textbf{If} there exists a feasible solution \textbf{then}\\
\quad \ \ \quad \quad $\mathcal{A}_{acc}=\mathcal{A}_{acc}\!\cup\!\{i^*\}$, $\mathcal{A}_{cur}\!=\!\mathcal{A}_{cur}\setminus\{i^*\}$,\\
\quad \ \ \quad \quad $\hat{\Theta}=\hat{\Theta}\cup\bar{\theta}_{i^*j}^l$, \textbf{break};\\
\quad \ \ \quad \textbf{Else}\\
\quad \ \ \quad \quad Recover the value of $\bar{\theta}_{i^*j}^l$ and its associated $\{\bar{\theta}_{i^*j'}^{l'}\}$;\\
\quad \ \ \quad \quad $\bar{\Theta}=\bar{\Theta}\setminus\{\bar{\theta}_{i^*j}^l\}$;\\
\quad \ \ \textbf{For} {each $i\in\mathcal{A}_{cur}$, where $\bar{\theta}_{ij}^l\notin\bar{\Theta}, \forall j, \forall l$} \textbf{do}\\{
\quad \ \ \quad \quad $\mathcal{A}_{rej}=\mathcal{A}_{rej}\!\cup\!\{i\}$, $\mathcal{A}_{cur}\!=\!\mathcal{A}_{cur}\setminus\{i\}$;\\
}
}
}
}
\end{algorithm}

We respectively introduce $\mathcal{A}_{acc}$ to indicate the set of chosen clients, $\mathcal{A}_{rej}$ to indicate the set of rejected clients and $\mathcal{A}_{cur}$ to indicate the set of undecided clients. Initially, $\mathcal{A}_{acc}$ and $\mathcal{A}_{rej}$ are empty, and $\mathcal{A}_{cur}$ includes all the client $i\!\in\!\mathcal{N}$. We relax the integer control variables involving the client $i\in\mathcal{A}_{cur}$ and further exploit some standard algorithms such as simplex and interior-point method to solve the relaxed problem. Then, we sort the fractional solution $\bar{\bm{\theta}}$ in the descending order of value $\omega_{ij}^l\bar{\theta}_{ij}^l$, where $\omega_{ij}^l$ refers to $p_i\!+\!\lambda Q_i\!-\!\rho(\alpha_{ij}'\!+\!\sum\nolimits_e\beta_{e}'\mathbbm{1}_{ij}^{l(e)}\varphi_{ij}^*)$ in the objective. We denote by $\bar{\Theta}$ the ordered list of $\bar{\bm{\theta}}$. For each $\bar{\theta}_{i^*j}^l\in\bar{\Theta}$ in order, we directly round it up to 1 and the other variables involving client $i^*$ down to 0. To proceed, we resort to Satisfiability Modulo Theories (SMT) \cite{barrett2018satisfiability} to judge if there is a feasible solution under the three constraints in problem $\mathcal{P}_1$.
If there exists a feasible solution, then we set $\mathcal{A}_{acc}=\mathcal{A}_{acc}\cup\{i^*\}$, $\mathcal{A}_{cur}\!=\!\mathcal{A}_{cur}\setminus\{i^*\}$ and record the rounded control variable (i.e., $\hat{\Theta}=\hat{\Theta}\cup\bar{\theta}_{i^*j}^l$). Otherwise, the rounding of $\bar{\theta}_{i^*j}^l$ and the associated variables involving client $i^*$ will be undone. Besides, $\bar{\theta}_{i^*j}^l$ will be removed from $\bar{\Theta}$ and go back to consider the next ordered element in $\bar{\Theta}$. Note that if all the control variables involving client $i$ have been removed from $\bar{\Theta}$, which means client $i$ is ``incompatible" with those chosen clients, we therefore put it in the rejected set.

We repeat the preceding procedures until the undecided client set is empty. At that time, the derived solution for the problem $\mathcal{P}_1$ is saved by $\hat{\Theta}$, with which we can eventually derive the solution of the problem $\mathcal{P}_0$ with the parametric objective $\mathcal{P}_0(\bm{x},\bm{y},\rho)$ in terms of Theorem 1 and Corollary 1. If the difference of the parametric objective $\mathcal{P}_0(\bm{x},\bm{y},\rho)$ between two continuous iterations is within a given tolerance (i.e., convergence), then we derive the final solution for the original problem $\mathcal{P}_0$. Otherwise, we will proceed the problem solving with the updated $\rho$.

\textbf{Practical Discussions}. We next make some discussions regarding the algorithm performance and overhead.

\emph{Performance}: We believe \textsf{Refinery} can find a reasonably good solution for the original mixed-integer fractional programming problem $\mathcal{P}_0$, as the utilized two linearizing strategies (i.e., Dinkelbach's transform and Theorem 1) and the proposed greedy based rounding algorithm for the new variant of the Unsplittable Multi-commodity Flow Problem are the best that can be achieved to date. Although our approach cannot guarantee the theoretical performance, its performance is experimentally justified by extensive evaluations in the next section. Besides, as the \textsf{CPN-FedSL} controller does not require to make realtime multivariate scheduling, we consider the complexity of \textsf{Refinery}, mainly attributed to the operations to derive $\{\theta_{ij}^l\}$ (i.e., Alg. \ref{Alg1}), is acceptable. Briefly, as the size of $\mathcal{A}_{cur}$ is reduced by at least one in each ``while" iteration, the outer loop terminates in at most $N$ iterations. In addition, the inner loop iterates over
all available variables, which is at most $\mathcal{O}(NM|\mathcal{L}|)$. Moreover, the relaxed linearized problem and the SMT check can be solved in polynomial time. To sum up, we can derive $\{\theta_{ij}^l\}$ in polynomial time.

\emph{Overhead}: As mentioned in the \textbf{Remark} on page 3, in order to make multivariate scheduling for each training task, the \textsf{CPN-FedSL} controller will derive $q_k^C$, $q_k^S$ and $s_k$ in an offline manner, obtain $p_i$ during client registration, and collect $c_{it}$ periodically from client uploading. We consider the latter two operations are common behavior with acceptable overhead in the context of CPN \cite{ITU-CPN}. However, as the neural networks are getting deeper and deeper, a training model with several tens and even more layers is not surprising, which will lead to a high overhead to derive $q_k^C$, $q_k^S$ and $s_k$ per partition point $k$.

\begin{figure}[thb]
\centering
\setlength{\belowcaptionskip}{-0.3cm}
\subfigure[DenseNet.]{
\includegraphics[width=0.46\linewidth]{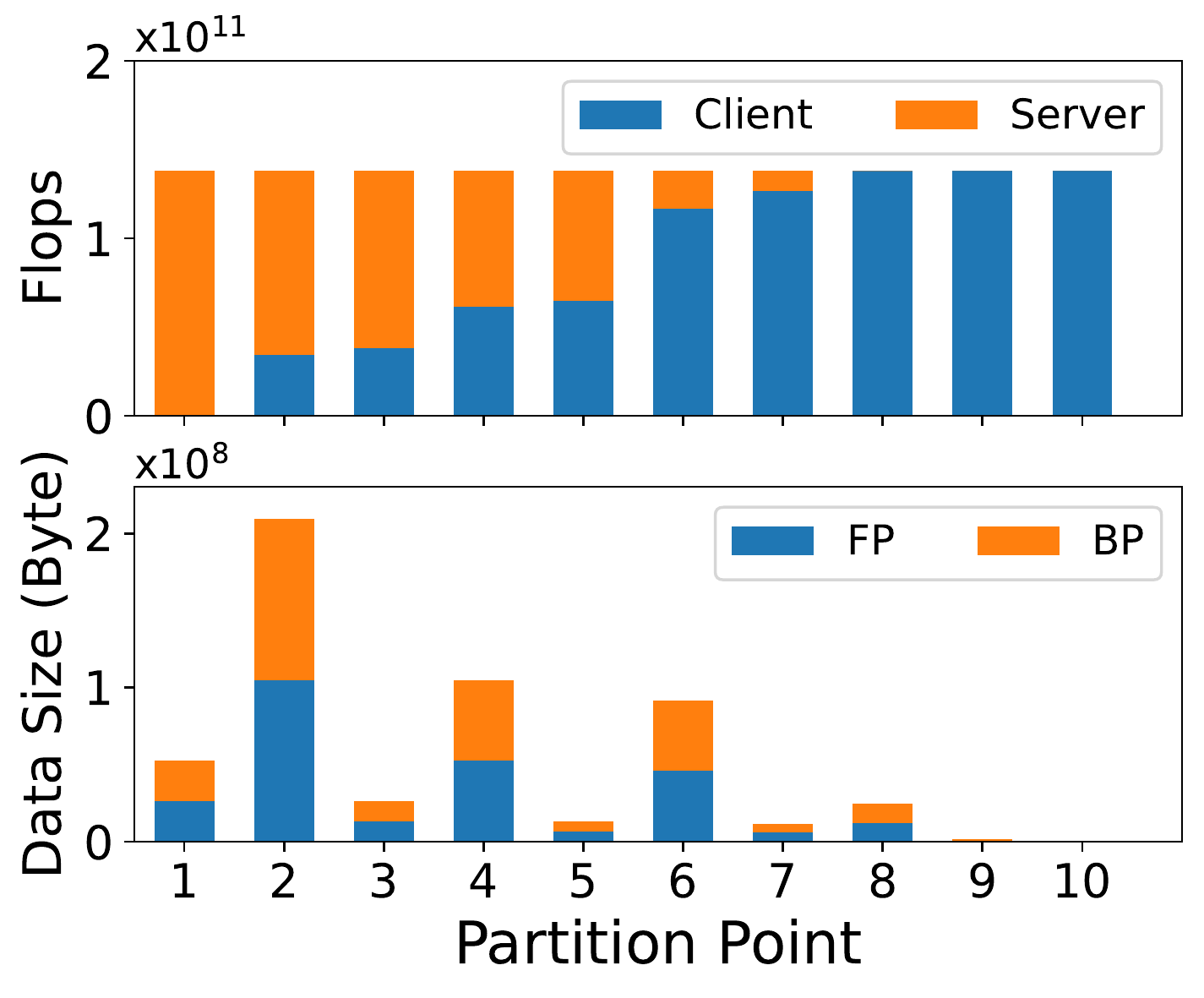}
}
\subfigure[MobileNet.]{
\includegraphics[width=0.46\linewidth]{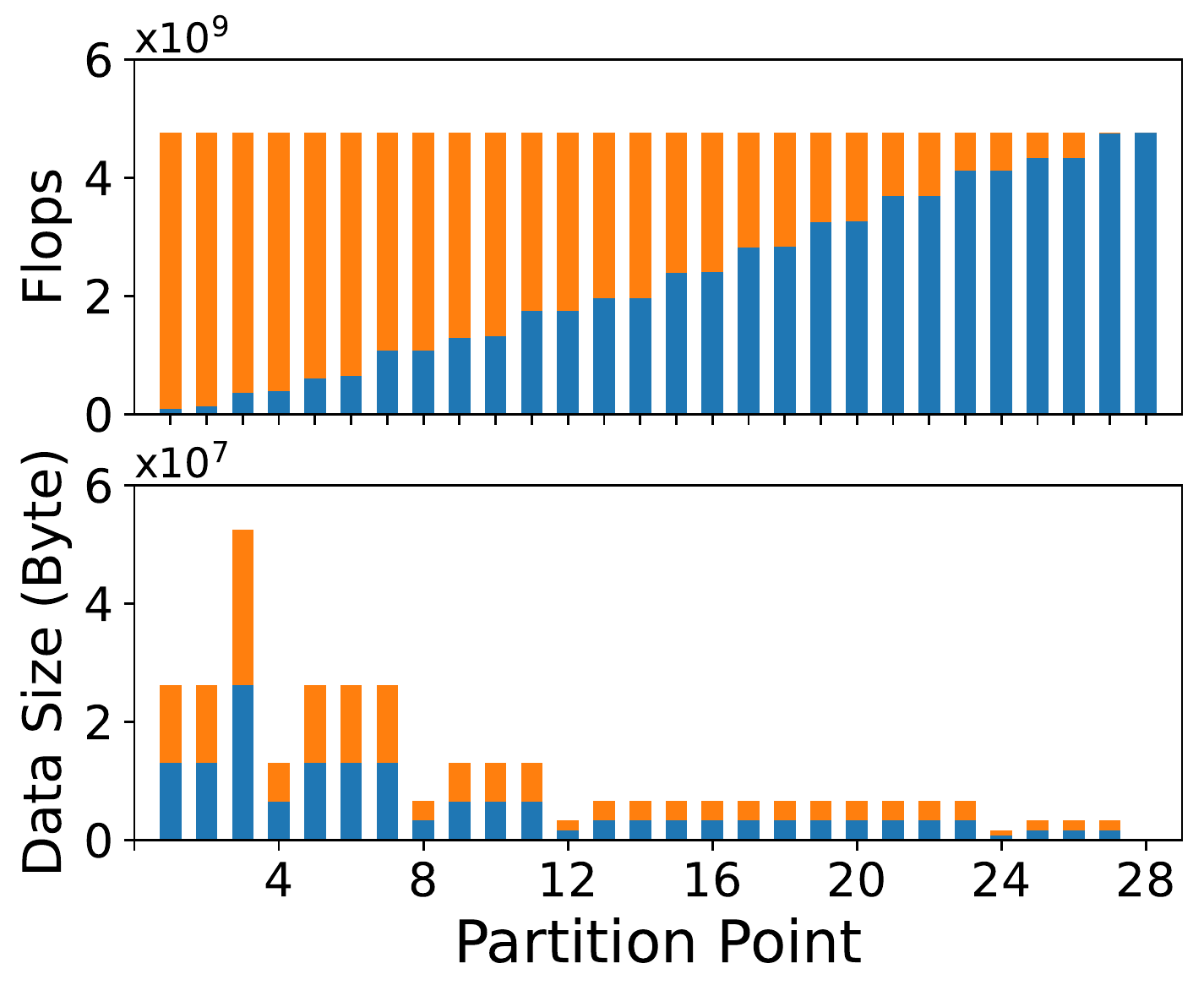}
}
\caption{The computing density and the amount of exchanged data of different training models (the number of trained data samples is 100).}\label{fig:models}
\end{figure}

Fortunately, we observe the number of ``effective" partition points is limited by investigating a variety of training models. Consider two kinds of training tasks: DenseNet \cite{huang2017densely}, a large model that includes 200 layers and 10 neural network modules\footnote{Note that it is unnecessary to partition a neural network module (i.e., including several intertwined layers), and thus we consider 10 partition points for DenseNet (i.e., partition between modules).}; MobileNet \cite{howard2017mobilenets}, a small model that includes 28 layers. Fig. \ref{fig:models} presents the corresponding computing density and the amount of exchanged data (i.e., FP activation and BP gradient). From the perspective of computing density, it is fairly regular that a later partition point corresponds to a higher computing density (i.e., more training layers) at the client side. However, there is no clear trend from the perspective of data exchange. In this context, we can check each partition point from the beginning to the end and filter out the ``effective" partition points. Specifically, we consider an effective partition point should satisfy \emph{the amount of data exchange at that point is much smaller than that at each pervious point, since a later partition point requires more computing capacity at the client side}. With this principle, we can obtain the effective partition points of DenseNet are 1, 3, 5, 9 and those of MobileNet are 1, 4, 8, 12, 24, which significantly contracts the partition point set and thus reduces the overhead to derive \{$q_k^C$, $q_k^S$, $s_k$\}.

\section{Performance Evaluation}\label{sec4}

In this section, we conduct extensive experiments to evaluate \textsf{CPN-FedSL} by answering the following questions.
\begin{itemize}
  \item What is the performance of \textsf{CPN-FedSL} compared with standard and SOTA learning frameworks? (\textbf{Exp\#1})
  \item Whether each part of multivariate control in \textsf{Refinery} has effect on the performance? (\textbf{Exp\#2})
  \item What is the performance of \textsf{Refinery} compared with \emph{de facto} heuristic methods? (\textbf{Exp\#3})
  \item How is the performance of the greedy based rounding algorithm in \textsf{Refinery}? (\textbf{Exp\#4})
\end{itemize}

\subsection{Experiment Settings}

\textbf{Training Task}. We consider two training tasks: DenseNet and MobileNet with the widely used dataset ImageNet \cite{deng2009imagenet}. The computing density and the amount of exchanged data of these two tasks have provided in Fig. \ref{fig:models}. The number of epoch is set to 1. To compromise with the following computing capacity setting at both client and server side as well as the computing density of each training model, the batch size is set to 8 for DenseNet and 4 for MobileNet. In addition, the time length of global round (i.e., the given deadline $\Delta$) is set to 150s for DenseNet and 5s for MobileNet.

\begin{figure}[thb]
\centering
\setlength{\belowcaptionskip}{-0.3cm}
{
\subfigure[NSFNET.]{
\includegraphics[width=0.225\textwidth]{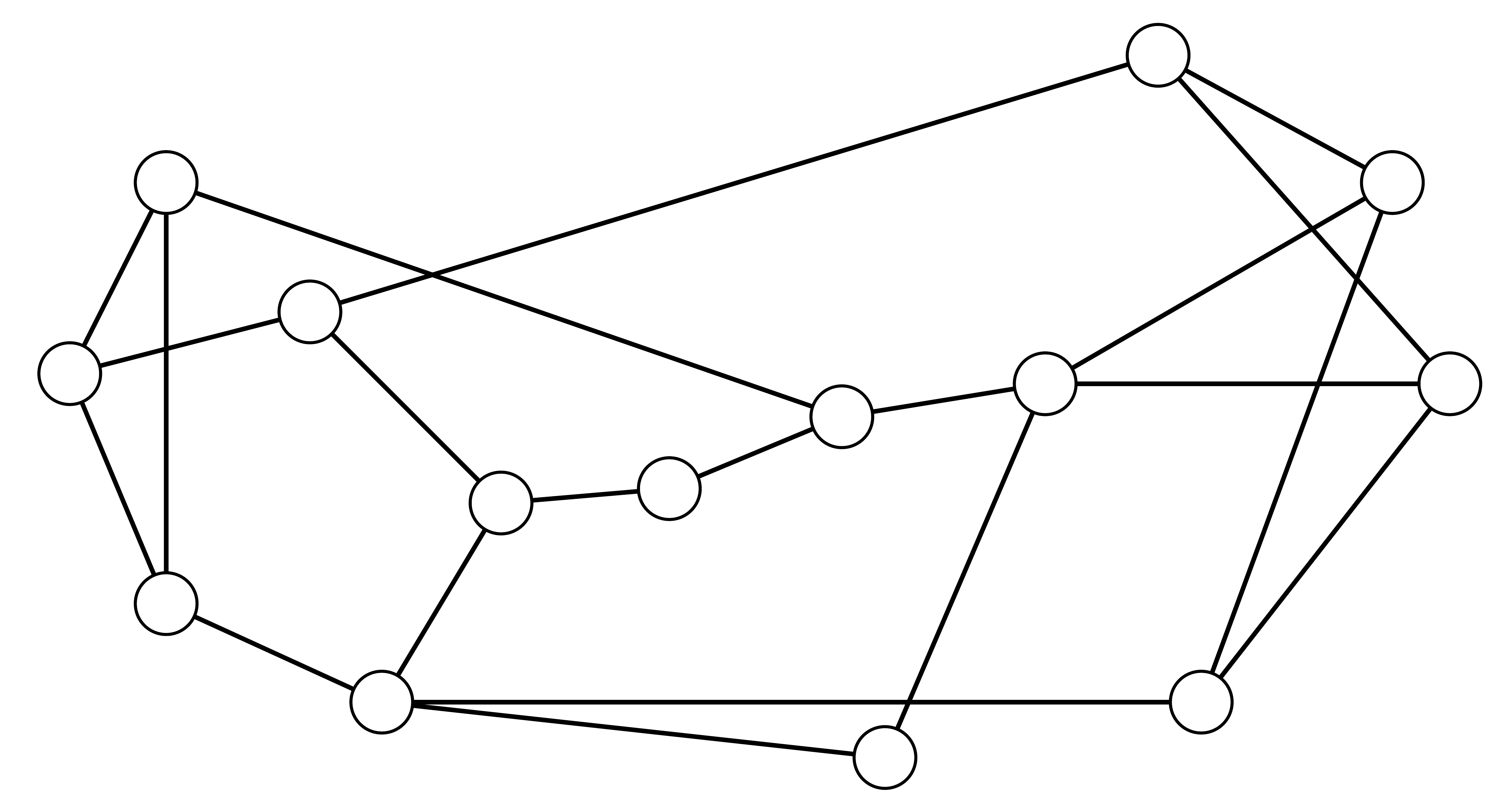}}
\subfigure[USNET.]{
\includegraphics[width=0.225\textwidth]{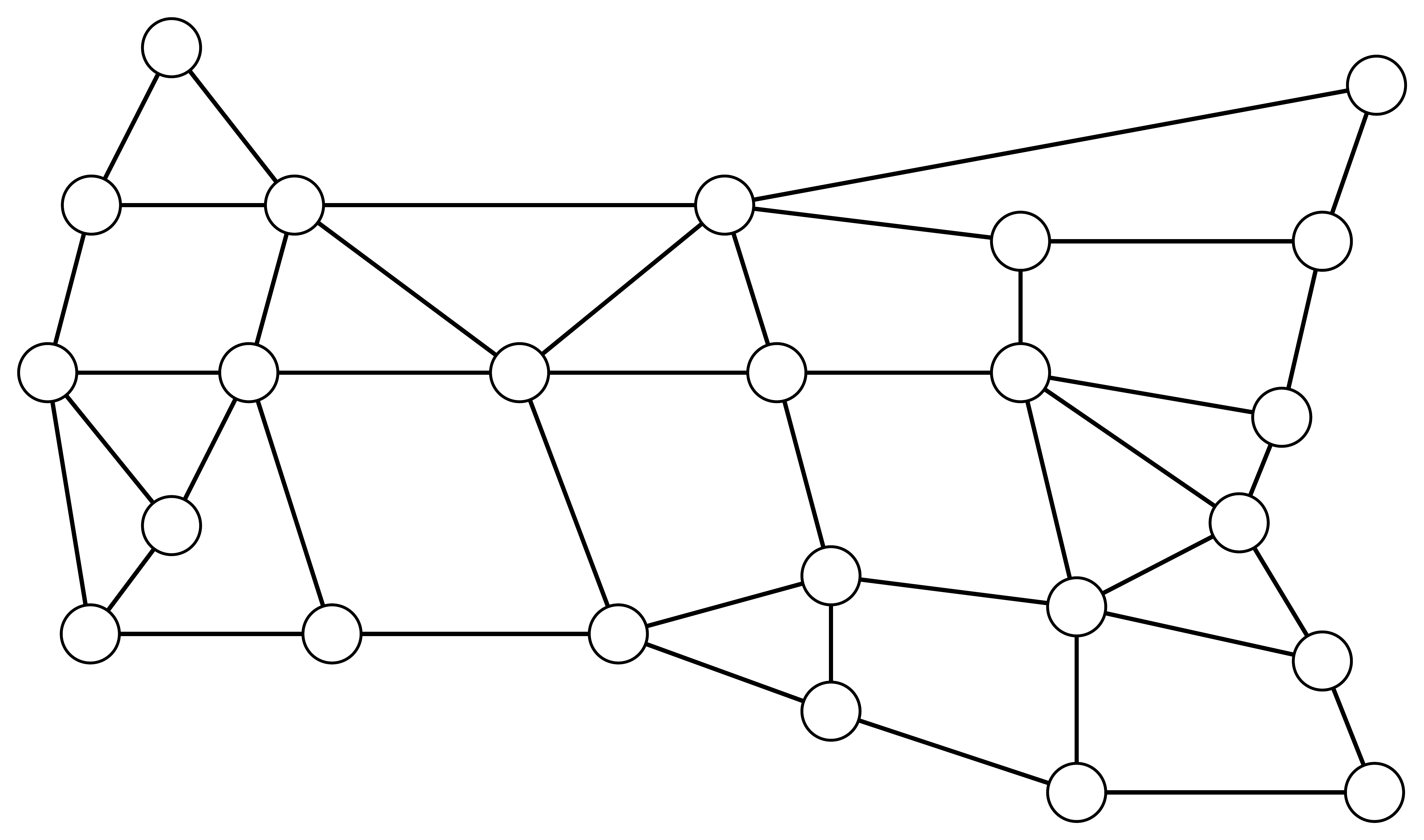}}
}
%\vspace{-0.2cm}
\caption{Network topology.}\label{fig:network}
%\vspace{-0.4cm}
\end{figure}

\textbf{Network Scenario}. We synthesize different scales of network scenarios for \textsf{CPN-FedSL}, in terms of two realistic network topologies as shown in Fig. \ref{fig:network}, where NSFNET stands for a small topology including 14 nodes and 21 links, and USNET stands for a more dense topology including 24 nodes and 43 links. We consider four network settings with different client distributions:
\begin{itemize}
  \item \textbf{NS1}: randomly select 6 nodes as computing sites and the rest 8 nodes each connects with 6 clients in NSFNET;
  \item \textbf{NS2}: randomly select 6 nodes as computing sites and 16 nodes each connecting with 1 client in USNET;
  \item \textbf{NS3}: randomly select 6 nodes as computing sites and 16 nodes each connecting with 3 clients in USNET;
  \item \textbf{NS4}: randomly select 6 nodes as computing sites and 3 nodes each connecting with 16 clients in USNET.
\end{itemize}

\vspace{-0.3cm}
\begin{table}[thb]
\centering
\caption{The settings of computing sites.}\label{tlb1}
\begin{tabular}{|c|c|c|c|c|c|c|}
  \hline
   & Site 1 & Site 2 & Site 3 & Site 4 & Site 5 & Site 6 \\ \hline
  Capacity & 4400 & 4400 & 4400 & 6500  & 6500  & 6500    \\ \hline
  Utilization & 5\% & 10\% & 15\%  & 5\% & 10\% & 15\% \\
  \hline
  Cost & 800 & 800 & 800  & 1500 & 1500 & 1500 \\
  \hline
\end{tabular}
\end{table}

The settings of 6 computing sites are given in Tab. \ref{tlb1}, and the actual computing capacity is ``Capacity$\times$Utilization". The number of training servers per computing site is set to 8 for NS1, NS3 and NS4 while 3 for NS2. The unit bandwidth cost of network link is randomly from 1 to 10 for DenseNet and from 0.1 to 1 for MobileNet. The bandwidth capacity of network link is randomly from 3000 to 5000.

\textbf{Data Client}. We consider three kinds of client computing capacity: 400, 800 and 1200. Each client is assigned a capacity with a random utilization from 2\% to 20\% per global round. Each client has a random number of data samples from 4000 to 20000. The client weight $p$ is calculated in terms of the size of dataset (i.e., $p_i=|D_i|/\sum_{i=1}^N|D_i|$). Besides, we introduce an auxiliary parameter $p'\!=10000\!$ to balance the value of training utility and system cost (i.e., $p_i\!=\!p_ip'$).

\subsection{Experiment Evaluation}

\textbf{(Exp\#1) Different Learning Frameworks}. We consider the following distributed
machine learning frameworks.
\begin{itemize}
  \item \textbf{FedAvg} \cite{li2019convergence}: each client will conduct the local training if it can accomplish it within the given deadline, which is the lower bound of learning performance;
  \item \textbf{SplitFed (Unlimited)} \cite{thapa2020splitfed}: select the computing site with the largest computing capacity and choose the partition point to benefit the most of clients without the constraints on link bandwidth and the number of training servers, which is the upper bound of learning performance;
  \item \textbf{SplitFed (Limited)}: similar to SplitFed (Unlimited) while taking two capacity constraints into account;
  \item \textbf{CPN-FedSL (NQ)}:   similar to \textsf{CPN-FedSL} without taking the fairness-aware client admission into account.
\end{itemize}

We create Python programs to conduct the performance comparison under each network scenario (i.e., NS1\! --\! NS4) via simulation. For ease of evaluation, we consider each experiment runs 30 training rounds%\footnote{Despite different settings have impact on the model accuracy of each learning framework, such as increasing the number of training round can contribute to speeding up model convergence for any learning frameworks, they do change the trend of framework performance comparison.}
(i.e., $T\!=\!30$), and exploit \emph{Average Training Amount}, the average number of trained data samples per round as an approximate metric to indicate the convergence time (due to the simulation based evaluation), since training more data samples per round is beneficial for model convergence to some extent \cite{li2019convergence}. In addition, we consider another performance metric \emph{Normalized Accuracy}, the derived accuracy of each distributed learning framework with its trained data samples compared with the centralized learning with all the data samples.

Note that, in order to derive the normalized accuracy in practice, we exploit three desktops to respectively act as client, training server and parameter server. As each pair of client and server trains the model in parallel, we exploit the real pair to mimic the behavior of each virtual pair in the experiment. That is, the real pair sequentially trains the model in terms of the decided partition point for each virtual pair per round, while uploading the derived model parameters of each virtual pair to the parameter server. The parameter server conducts the FedAvg operation \cite{li2019convergence} when receiving the model parameters from all the virtual pairs of client and server.

\begin{table*}[t]
	\centering
	\caption{Performance comparison among different learning frameworks (DenseNet and MobileNet).}
	\begin{tabular}{|c|c|c|c|c|c|c|c|c|c|c|c|c|c|c|c|c|}
		\hline
		
		\multirow{3}{*}{\begin{tabular}[c]{@{}c@{}}Learning\\ Framework\end{tabular} } & \multicolumn{8}{c|}{DenseNet} & \multicolumn{8}{c|}{MobileNet} \\ \cline{2-17}
		&\multicolumn{4}{c|}{ Training Amount (x$10^4$)} & \multicolumn{4}{c|}{Normalized Accuracy (\%)} & \multicolumn{4}{c|}{ Training Amount (x$10^4$)} & \multicolumn{4}{c|}{Normalized Accuracy (\%)} \\ %\multicolumn{4}{*}{CIFAR-10}&\multicolumn{4}{*}{SUN}\\
		\cline{2-17}
		
		& NS1 & NS2 & NS3 & NS4 & NS1 & NS2 & NS3 & NS4 & NS1 & NS2 & NS3 & NS4 & NS1 & NS2 & NS3 & NS4 \\
		\hline
		
		%0.25x-----------------
		FedAvg & 13.7 &5.3 & 13.1& 12.9 & 81.6 & 72.3 & 80.6 & 81.1 & 14.1&5.5 & 14.9& 14.1 & 60.5 & 56.7 & 61.2 & 60.7 \\
		\hline
		
		SplitFed (Unlimited)& 58.9 & 19.3 & 59.2 & 59.2& 95.6 & 84.3 & 96.1 & 96.1 & 60.1 & 20.2 & 60.8 & 60.8 & 89.7 & 69.4 & 89.9 & 89.9 \\
		\hline
		
		SplitFed (Limited)& 24.6& 8.7 & 25.2 & 25.1 &87.7 &80.1 & 87.9 & 87.9 & 24.5 & 8.7 & 25.7 & 25.2 &75.6 &58 & 77.1 & 76.9 \\
		\hline
		
		CPN-FedSL (NQ) & 44.6 &15.1&48.5 &51.0 &94.3 &82.3 &94.9 &95.2 & 44.6 & 11.8& 48.9 & 46.8 &84.5 &59.2 &85.1 &83.6 \\
		\hline
		
		\textbf{CPN-FedSL} & \textbf{42.9} &\textbf{13.5} &\textbf{48.5} &\textbf{46.1} &\textbf{95.3} &\textbf{82.8}& \textbf{96.1} & \textbf{95.8} & \textbf{42.2} & \textbf{13.3} & \textbf{42.2} & \textbf{42.1} & \textbf{87.0} & \textbf{61.5} &\textbf{86.9} &\textbf{86.5} \\
		\hline
	\end{tabular}
	\label{tab2}
\end{table*}

\begin{figure*}[tt]
\setlength{\belowcaptionskip}{-0.5cm}
\begin{minipage}{0.5\linewidth}
\centering
{
\subfigure[\small DenseNet.]{
\includegraphics[width=0.45\linewidth]{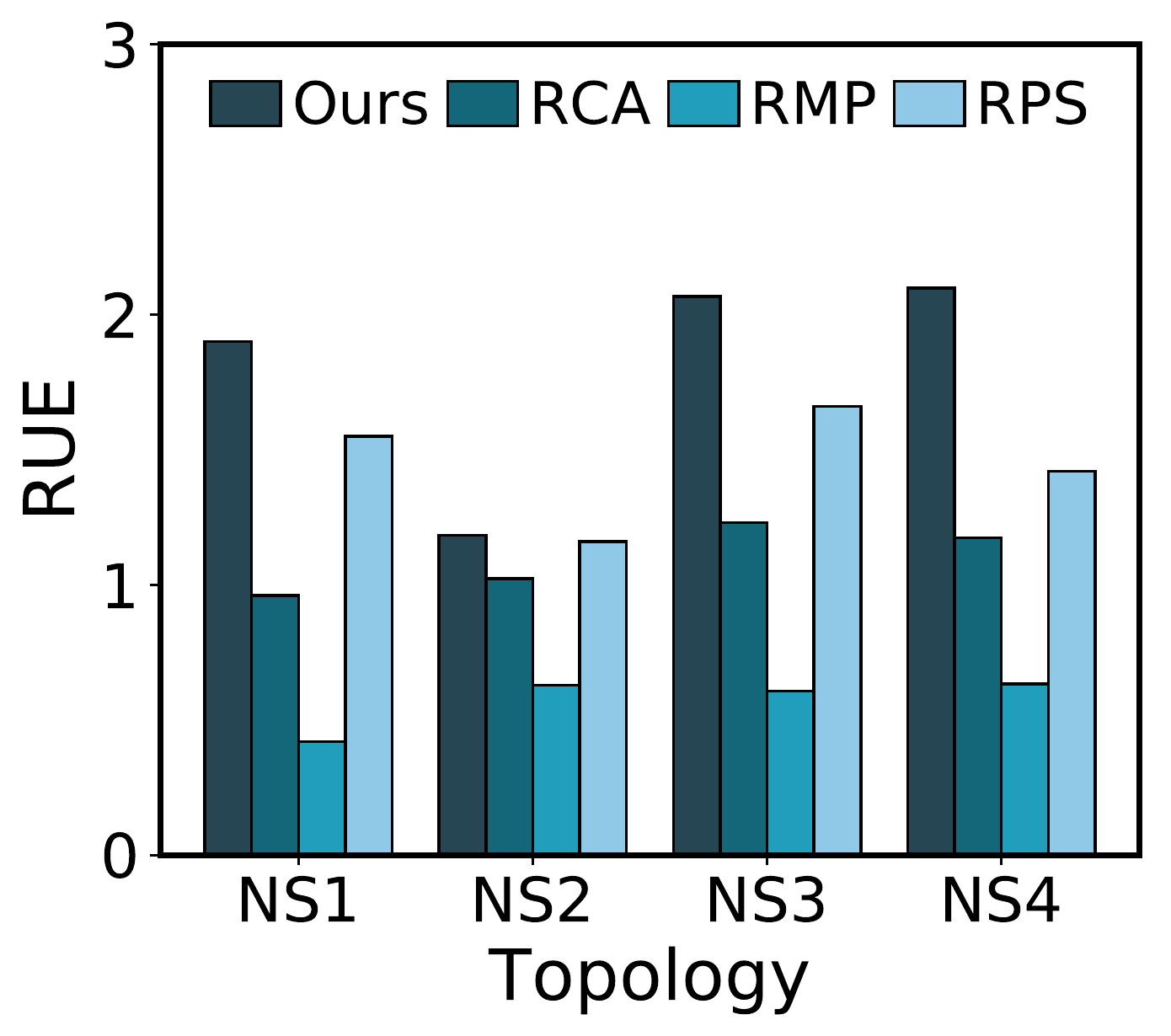}}
\
\subfigure[\small MobileNet.]{
\includegraphics[width=0.45\linewidth]{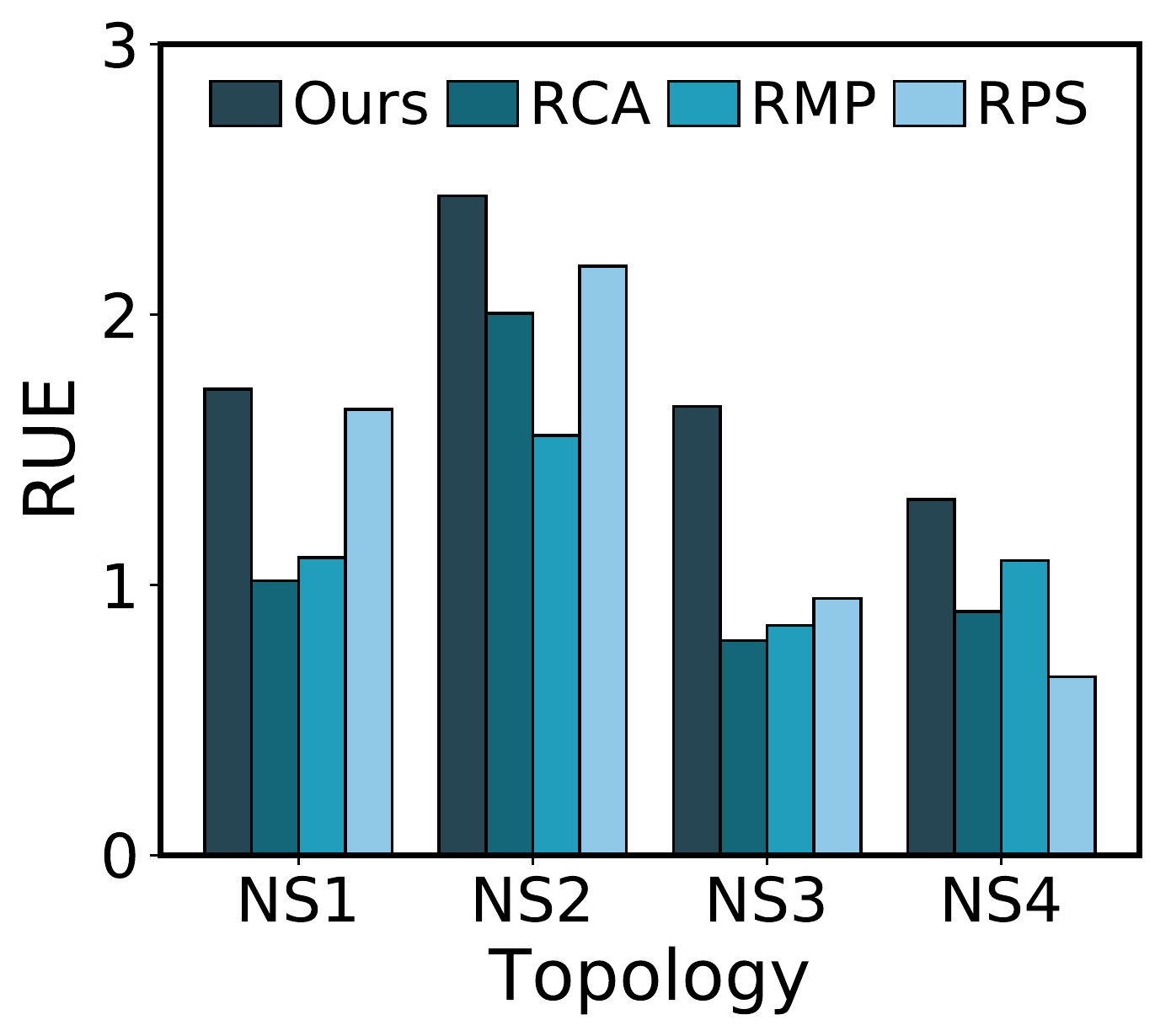}}}
\caption{Performance comparison among different variants.}
\label{fig:part}
\end{minipage}
\begin{minipage}{0.5\linewidth}
\centering
{
\subfigure[\small DenseNet.]{
\includegraphics[width=0.45\linewidth]{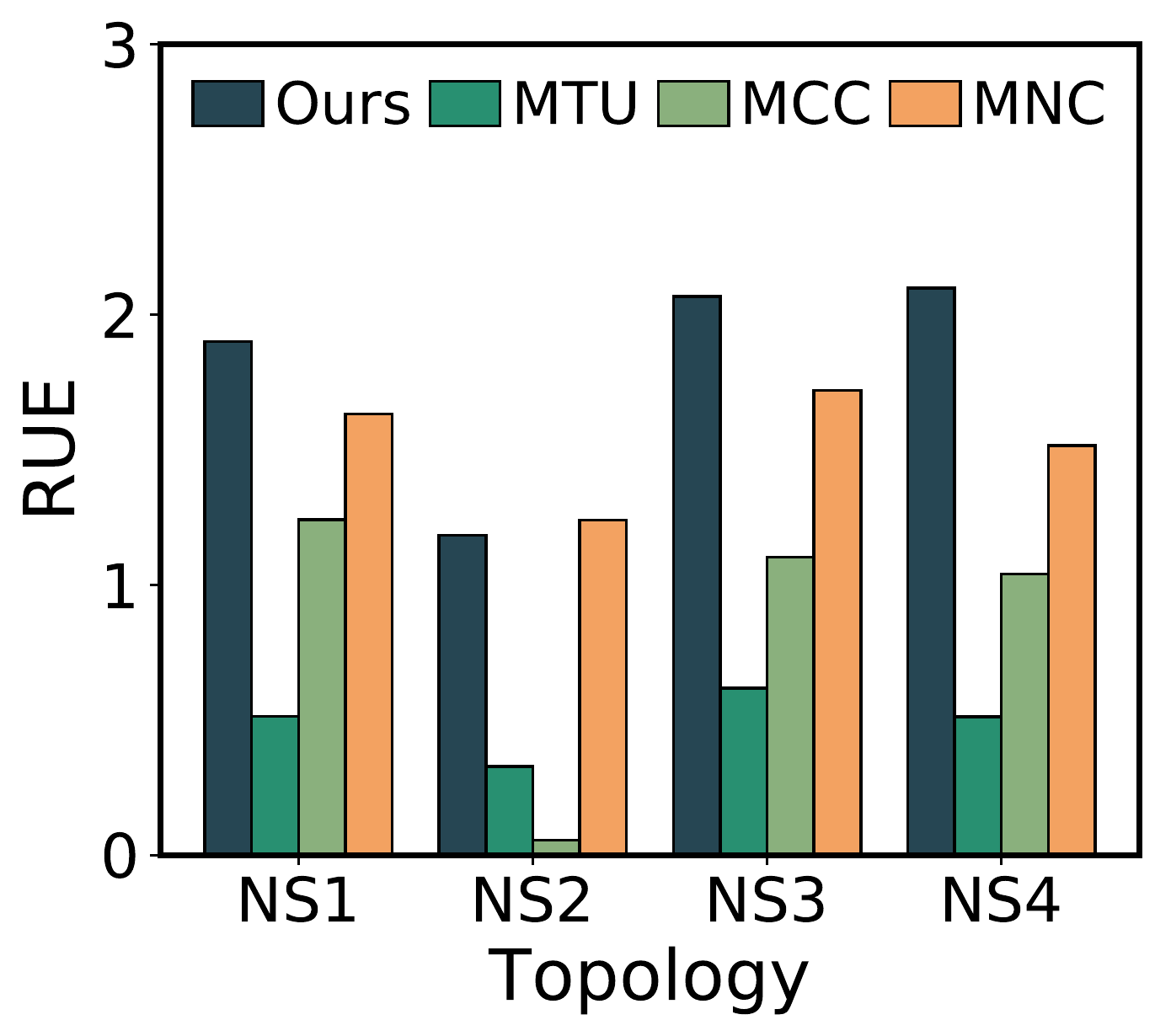}}
\
\subfigure[\small MobileNet.]{
\includegraphics[width=0.45\linewidth]{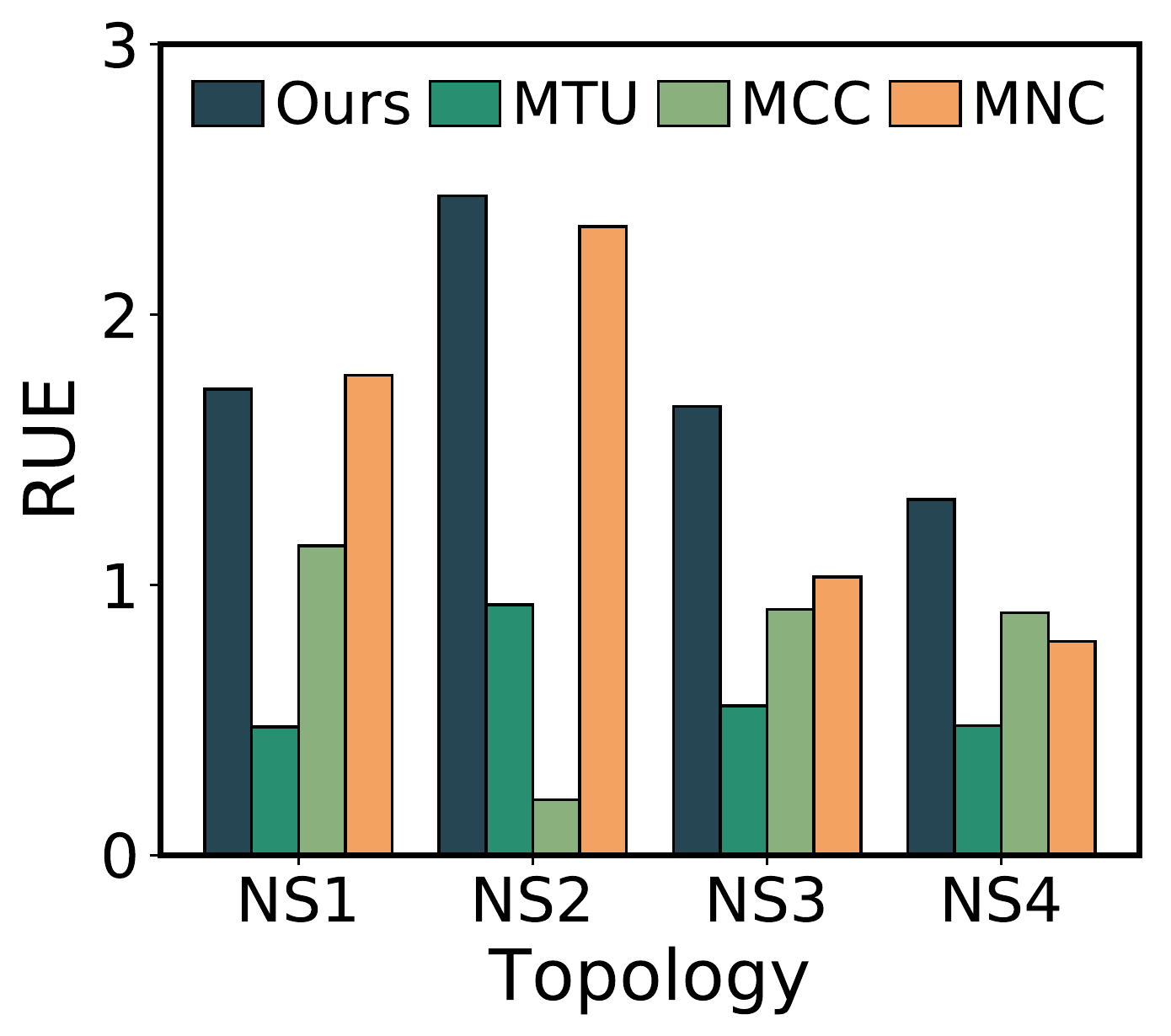}}}
\caption{Performance comparison among different heuristic methods.}
\label{fig:greedy}
\end{minipage}
%\vspace{-0.4cm}
\end{figure*}

The evaluation results are presented in Tab. \ref{tab2}. Overall, as for each training task, the performance of each SL enabled framework significantly outperforms that of FedAvg in each network scenario. These results indicate the great potential by integrating FL with SL (i.e., more computing resources). Our performance (in bold) can achieve the closest performance (i.e., normalized accuracy) compared with SplitFed (Unlimited), and can increase the normalized accuracy by 11.4\% (8.2\%) compared with SplitFed (Limited) for MobileNet (DenseNet). These results reveal the benefits of adopting various servers and flexible partition for model training. Despite the average training amount of \textsf{CPN-FedSL} is slightly lower than that of CPN-FedSL (NQ), the accuracy on the contrary is slightly higher. This result emphasizes the effectiveness of fairness-aware client admission. To sum up, we can conclude that \textsf{CPN-FedSL} is superior to standard and state-of-the-art learning frameworks under a variety of settings.

\textbf{(Exp\#2) Different Variants of \textsf{Refinery}}. We consider the following three variants (i.e., ablation experiments).
\begin{itemize}
  \item \textbf{Replaced Client Admission (RCA)}: randomly choose each client in terms of the client weight $p$;
  \item \textbf{Replaced Model Partition (RMP)}: partition the model at the same point (i.e., only a single partition point);
  \item \textbf{Replaced Path Selection (RPS)}:  select the shortest path after the decision-making of server selection.
\end{itemize}

We exploit the same simulation setting with Exp\#1 except the evaluation metric is RUE, the integrated metric introduced in this paper. The evaluation results are presented in Fig. \ref{fig:part}. Compared with RCA, \textsf{Refinery} improves the average RUE by 1.84$\times$ in NS1, by 1.19$\times$ in NS2, by 1.89$\times$ in NS3 and by 1.62$\times$ in NS4. These results indicate that although RCA can also achieve the fairness-aware client admission, it could not fully utilize the available computing resources (i.e., some clients are not chosen due to the randomness). In other words, our client admission that jointly captures fairness and effectiveness is more preferable. Compared with RMP, \textsf{Refinery} improves the average RUE by 3.04$\times$ in NS1, by 1.73$\times$ in NS2, by 2.68$\times$ in NS3 and by 2.26$\times$ in NS4. These results emphasize that flexible model partition is more significant compared with the randomized client admission. Indeed, with the single partition point, some pairs of client and server may lead to more data exchange (e.g., not the optimal partition point), and therefore increase the system cost. Compared with RPS, \textsf{Refinery} improves the average RUE by 1.14$\times$ in NS1, by 1.07$\times$ in NS2, by 1.50$\times$ in NS3 and by 1.74$\times$ in NS4. We can observe that RPS is good at the sparse scenario (i.e., NS2) while weak in the dense scenario (i.e., NS4). As \textsf{Refinery} jointly takes server selection and routing into account, its RUE is always greater than 1. In addition, it can achieve better performance when the clients are well distributed in the network (i.e., NS3). To sum up, we can conclude that each part of the multivariate scheduling in \textsf{Refinery} has a significant effect on the RUE performance.

\textbf{(Exp\#3) Different Heuristic Methods}. We consider the following three \emph{de facto} heuristic methods.
\begin{itemize}
  \item \textbf{Maximize Training Utility (MTU)}: sort the clients in ascending order of client computing capacity, and then select the computing site with the largest computing capacity for each client in order (If one site is fully filled, then select the second largest one, and so on);
  \item \textbf{Minimize Computing Cost (MCC)}: shuffle the clients, and then select the computing site with the lowest unit server cost for each client in order (If one site is fully filled, then select the second lowest one, and so on);
  \item \textbf{Minimize Network Cost (MNC)}: select the computing site for each client in terms of their routing hops.
\end{itemize}

We exploit the same simulation setting with Exp\#2. The evaluation results are presented in Fig. \ref{fig:greedy}. Compared with MTU, \textsf{Refinery} improves the average RUE by 3.67$\times$ in NS1, by 3.12$\times$ in NS2, by 3.17$\times$ in NS3 and by 3.42$\times$ in NS4. These results indicate that solely optimizing the training utility could suffer from a high system cost. Therefore, we should jointly optimize them. Compared with MCC, \textsf{Refinery} improves the average RUE by 1.51$\times$ in NS1, by 16.57$\times$ in NS2, by 1.85$\times$ in NS3 and by 1.74$\times$ in NS4. Remarkably, MCC is weak in the sparse scenario, since the random shuffle and cost-driven server selection could not always find suitable pairs of client and server, and accordingly lead to a relatively low training utility (i.e., a low RUE). Compared with MNC, \textsf{Refinery} improves the average RUE by 1.07$\times$ in NS1, by 1.01$\times$ in NS2, by 1.41$\times$ in NS3 and by 1.52$\times$ in NS4. We can observe that MNC is pretty good at the small-scale and sparse scenario (i.e., NS1 and NS2), and can achieve a slightly better performance than ours in some cases. This is because when the scenario is sparse, the distance-driven server selection could not interference with each other. Indeed, when the network scenario becomes denser (i.e., NS2$\to$NS1$\to$NS3$\to$NS4), the performance of MNC cannot catch up with ours. To sum up, we can conclude that \textsf{Refinery} significantly outperforms \emph{de facto} heuristic methods in most cases.

\begin{figure}[thb]
\centering
\subfigure[DenseNet.]{
\centering\includegraphics[height=1.4in]{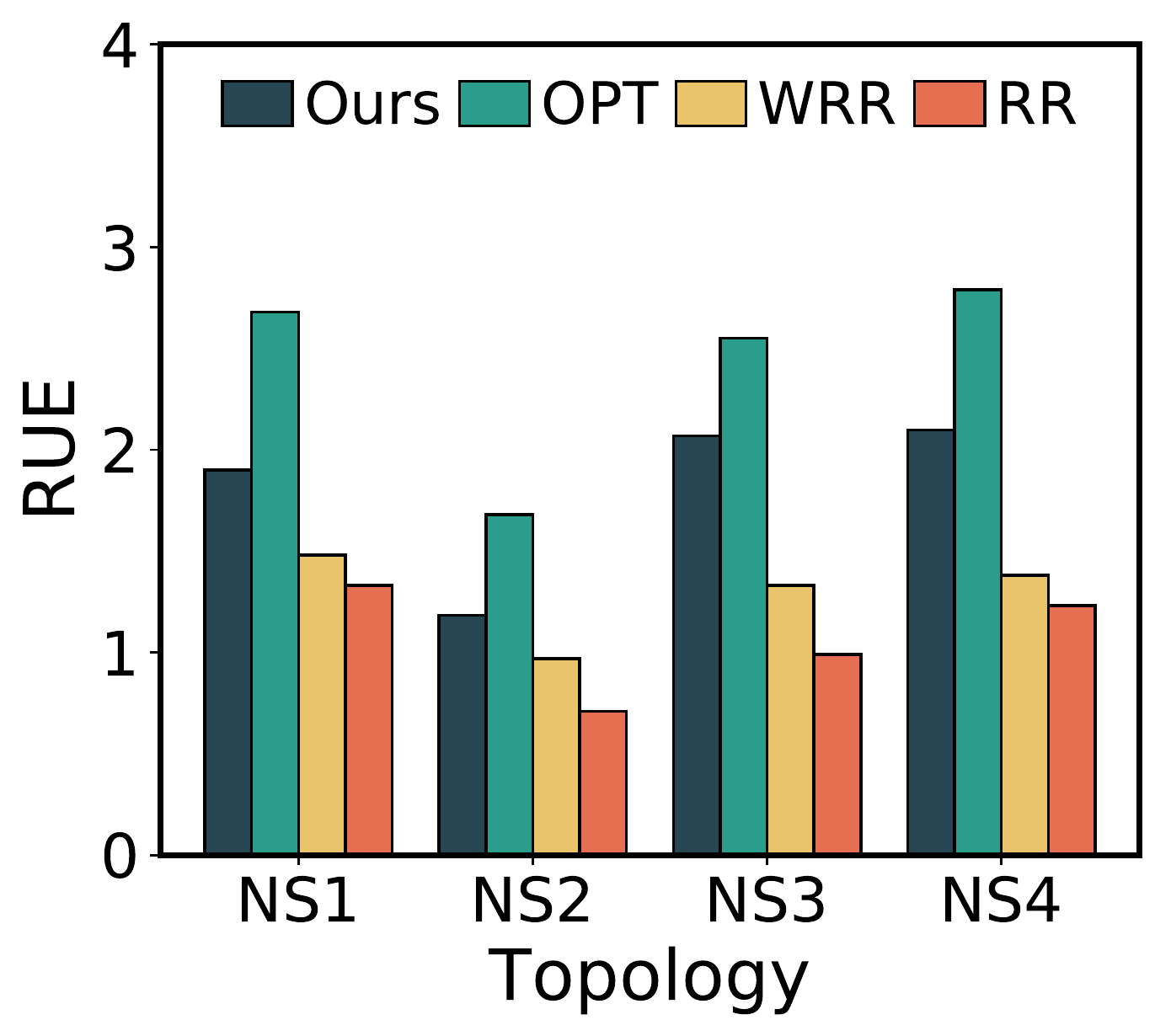}
}
\subfigure[MobileNet.]{
\centering\includegraphics[height=1.4in]{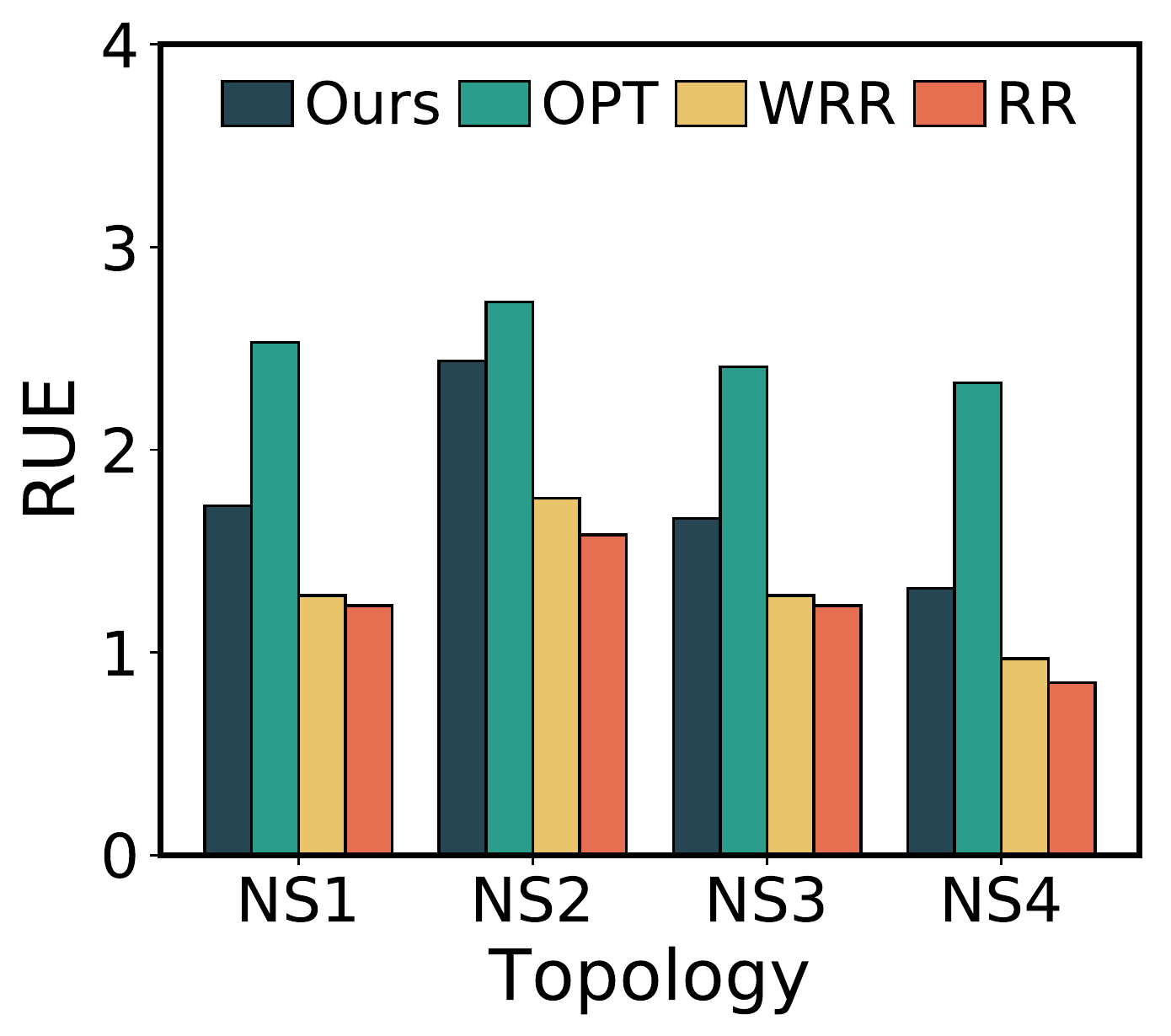}
}
\caption{Performance comparison among different algorithms for $\mathcal{P}_1$.}
\label{fig:round}
\vspace{-0.2cm}
\end{figure}

\textbf{(Exp\#4) Different Algorithms for $\mathcal{P}_1$}. We consider the following optimal solution and two alternative algorithms.
\begin{itemize}
  \item \textbf{Optimal Solution (OPT)}: directly solve the integer programming $\mathcal{P}_1$ via the GLPK solver;
  \item \textbf{Weighted Randomized Rounding (WRR)}: solve the relaxed problem and conduct the randomized rounding in terms of $\omega_{ij}^l\bar{\theta}_{ij}^l$;
  \item \textbf{Randomized Rounding (RR)}: solve the relaxed problem and conduct the randomized rounding in terms of $\bar{\theta}_{ij}^l$.
\end{itemize}

We exploit the same simulation setting with Exp\#2 (That is, we substitute our proposed algorithm with them in \textsf{Refinery} to evaluate the performance). The evaluation results are presented in Fig. \ref{fig:round}. Overall, we can observe that the performance of WRR is better than that of RR, since the former algorithm fully takes the objective into account. Compared with WRR, our greedy based rounding algorithm improves the average RUE by 1.32$\times$ in NS1, by 1.30$\times$ in NS2, by 1.43$\times$ in NS3 and by 1.44$\times$ in NS4. This is because WRR can be roughly viewed as a one-time operation in the outer loop of ours without the SMT check. Compared with OPT, our performance will account for 68\% in NS1, 80\% in NS2, 75\% in NS3 and 66\% in NS4 (i.e., 72.8\% on average). To sum up, we can conclude that the proposed greedy based rounding algorithm is reasonably good compared with alternative algorithms.

\section{Conclusion}
In this paper, we advocate \textsf{CPN-FedSL}, a novel federated split learning framework over computing power network. We build a dedicated model to capture its basic settings and learning characteristics. Based on this model, we introduce a novel performance metric integrating training utility with system cost and formulate a multivariate scheduling problem. We design \textsf{Refinery}, an efficient approach that first linearizes the fractional objective and non-convex constraints, and then solves the transformed problem via a greedy based rounding algorithm in multiple iterations. Extensive evaluations corroborate the superior performance of \textsf{CPN-FedSL} and \textsf{Refinery}.

\section{Acknowledgment}

We thank the anonymous reviewers for their constructive feedback. This work was supported in part by the National Natural Science Foundation of China (No. 62172241), the Natural Science Foundation of Tianjin (No. 20JCZDJC00610) and the U.S. National Science Foundation (CNS-2047719 and CNS-2225949).

\bibliographystyle{ieeetr}
\bibliography{iwqos2023}

\begin{thebibliography}{10}

\bibitem{ITU2030}
``{ITU-T Technical Report: Representative use cases and key network
  requirements for Network 2030}.'' Available in:
  \url{https://www.itu.int/dms_pub/itu-t/opb/fg/T-FG-NET2030-2020-SUB.G1-PDF-E.pdf}.

\bibitem{huawei2030}
``{Huawei Industry Report: Communications Network 2030}.'' Available in:
  \url{https://www-file.huawei.com/-/media/corp2020/pdf/giv/industry-reports/communications_network_2030_en.pdf}.

\bibitem{krol2019compute}
M.~Kr{\'o}l, S.~Mastorakis, {\em et~al.}, ``{Compute first networking:
  Distributed computing meets icn},'' in {\em ACM ICN}, 2019.

\bibitem{crowcroft2021compute}
J.~Crowcroft, P.~Eardley, {\em et~al.}, ``{Findings and Recommendations for
  Compute First Networking},'' in {\em Dagstuhl Reports}, 2021.

\bibitem{ITU-CPN}
``{ITU-T Technical Report: Computing power network ---– Framework and
  architecture}.'' Available in:
  \url{https://www.itu.int/rec/T-REC-Y.2501-202109-I}.

\bibitem{stoica2021cloud}
I.~Stoica and S.~Shenker, ``From cloud computing to sky computing,'' in {\em
  ACM HotOS}, 2021.

\bibitem{thapa2020splitfed}
C.~Thapa, M.~A.~P. Chamikara, {\em et~al.}, ``{Splitfed: When federated
  learning meets split learning},'' in {\em AAAI}, 2022.

\bibitem{han2021accelerating}
D.-J. Han, H.~I. Bhatti, {\em et~al.}, ``Accelerating federated learning with
  split learning on locally generated losses,'' in {\em ICML}, 2021.

\bibitem{oh2022locfedmix}
S.~Oh, J.~Park, {\em et~al.}, ``{LocFedMix-SL: Localize, Federate, and Mix for
  Improved Scalability, Convergence, and Latency in Split Learning},'' in {\em
  ACM WWW}, 2022.

\bibitem{hongefficient}
J.~Hong, H.~Wang, {\em et~al.}, ``Efficient split-mix federated learning for
  on-demand and in-situ customization,'' in {\em ICLR}, 2022.

\bibitem{zhang2023privacy}
Z.~Zhang, A.~Pinto, {\em et~al.}, ``Privacy and efficiency of communications in
  federated split learning,'' {\em arXiv preprint arXiv:2301.01824}, 2023.

\bibitem{li2019convergence}
X.~Li, K.~Huang, {\em et~al.}, ``{On the Convergence of FedAvg on Non-IID
  Data},'' in {\em ICLR}, 2020.

\bibitem{yuan2020federated}
H.~Yuan and T.~Ma, ``{Federated accelerated stochastic gradient descent},'' in
  {\em NeurIPS}, 2020.

\bibitem{vepakomma2018split}
P.~Vepakomma, O.~Gupta, {\em et~al.}, ``{Split learning for health: Distributed
  deep learning without sharing raw patient data},'' {\em arXiv preprint
  arXiv:1812.00564}, 2018.

\bibitem{tuli2022splitplace}
S.~Tuli, G.~Casale, and N.~R. Jennings, ``{SplitPlace: AI Augmented Splitting
  and Placement of Large-Scale Neural Networks in Mobile Edge Environments},''
  {\em IEEE TMC}, 2022.

\bibitem{lim2020federated}
W.~Y.~B. Lim, N.~C. Luong, {\em et~al.}, ``Federated learning in mobile edge
  networks: A comprehensive survey,'' {\em IEEE Commun. Surv. Tut.}, 2020.

\bibitem{khan2021federated}
L.~U. Khan, W.~Saad, {\em et~al.}, ``Federated learning for internet of things:
  Recent advances, taxonomy, and open challenges,'' {\em IEEE Commun. Surv.
  Tut.}, 2021.

\bibitem{wu2022split}
W.~Wu {\em et~al.}, ``{Split Learning over Wireless Networks: Parallel Design
  and Resource Management},'' {\em arXiv preprint arXiv:2204.08119}, 2022.

\bibitem{krouka2021communication}
M.~Krouka, A.~Elgabli, {\em et~al.}, ``Communication-efficient split learning
  based on analog communication and over the air aggregation,'' in {\em IEEE
  GLOBECOM}, 2021.

\bibitem{gao2021evaluation}
Y.~Gao, M.~Kim, {\em et~al.}, ``{Evaluation and optimization of distributed
  machine learning techniques for internet of things},'' {\em IEEE TC}, 2021.

\bibitem{mao2017survey}
Y.~Mao, C.~You, {\em et~al.}, ``A survey on mobile edge computing: The
  communication perspective,'' {\em IEEE Commun. Surv. Tut.}, 2017.

\bibitem{sonkoly2021survey}
B.~Sonkoly, J.~Czentye, {\em et~al.}, ``Survey on placement methods in the edge
  and beyond,'' {\em IEEE Commun. Surv. Tut.}, 2021.

\bibitem{gao2021ocdst}
G.~Gao, L.~Wu, {\em et~al.}, ``Ocdst: Offloading chained dnns for streaming
  tasks,'' in {\em IEEE GLOBECOM}, 2021.

\bibitem{gao2019deep}
M.~Gao, W.~Cui, {\em et~al.}, ``Deep neural network task partitioning and
  offloading for mobile edge computing,'' in {\em IEEE GLOBECOM}, 2019.

\bibitem{he2020joint}
W.~He {\em et~al.}, ``Joint dnn partition deployment and resource allocation
  for delay-sensitive deep learning inference in iot,'' {\em IEEE IOTJ}, 2020.

\bibitem{zhang2021autodidactic}
L.~Zhang, L.~Chen, and J.~Xu, ``Autodidactic neurosurgeon: Collaborative deep
  inference for mobile edge intelligence via online learning,'' in {\em ACM
  WWW}, 2021.

\bibitem{dinkelbach1967nonlinear}
W.~Dinkelbach, ``On nonlinear fractional programming,'' {\em INFORMS Management
  science}, 1967.

\bibitem{he2022enabling}
L.~He, S.~Wang, {\em et~al.}, ``Enabling application-aware traffic engineering
  in ipv6 networks,'' {\em IEEE Network}, 2022.

\bibitem{report}
``{O}nline {T}echnical {R}eport.''
  \url{https://www.dropbox.com/s/mwikm0ox0y1xaac/technical%20report.pdf?dl=0.}

\bibitem{neely2010stochastic}
M.~J. Neely, ``Stochastic network optimization with application to
  communication and queueing systems,'' {\em Synthesis Lectures on
  Communication Networks}, 2010.

\bibitem{chakrabarti2007approximation}
A.~Chakrabarti, C.~Chekuri, {\em et~al.}, ``Approximation algorithms for the
  unsplittable flow problem,'' {\em Springer Algorithmica}, 2007.

\bibitem{raghavan1988probabilistic}
P.~Raghavan, ``Probabilistic construction of deterministic algorithms:
  approximating packing integer programs,'' {\em Elsevier Journal of Computer
  and System Sciences}, 1988.

\bibitem{barrett2018satisfiability}
C.~Barrett and C.~Tinelli, ``Satisfiability modulo theories,'' in {\em Springer
  Handbook of model checking}, 2018.

\bibitem{huang2017densely}
G.~Huang, Z.~Liu, {\em et~al.}, ``Densely connected convolutional networks,''
  in {\em IEEE CVPR}, 2017.

\bibitem{howard2017mobilenets}
A.~G. Howard, M.~Zhu, {\em et~al.}, ``Mobilenets: Efficient convolutional
  neural networks for mobile vision applications,'' {\em arXiv preprint
  arXiv:1704.04861}, 2017.

\bibitem{deng2009imagenet}
J.~Deng, W.~Dong, {\em et~al.}, ``Imagenet: A large-scale hierarchical image
  database,'' in {\em IEEE CVPR}, 2009.

\end{thebibliography}
\end{document}